**Acceleration of $Fe^{3+}/Fe^{2+}$ cycle in garland-like MIL-101(Fe)/$MoS_2$ nanosheets to promote peroxymonosulfate activation for sulfamethoxazole degradation**

Ke Zhu[a], Wenlei Qin[a], Yaping Gan[c], Yizhe Huang[a], Zhiwei Jiang[a], Yuwen Chen[a], Xin Li[a], Kai Yan[a,b,*]

[a] Guangdong Provincial Key Laboratory of Environmental Pollution Control and Remediation Technology, School of Environmental Science and Engineering, Sun Yat-sen University, Guangzhou 510275, Guangdong, China

[b] Guangdong Laboratory for Lingnan Modern Agriculture, South China Agricultural University, Guangzhou 510642, Guangdong, China

[c] Ecology and Health Institute, Hangzhou Vocational&Technical College, Hangzhou 310018, China

* Corresponding author Email address: yank9@mail.sysu.edu.cn (K. Yan)

**Abstract**

Iron-based molybdenum disulfide (Fe-$MoS_2$) has emerged as a Fenton-like catalyst for the highly efficient degradation of antibiotics, but the structure-activity relationship remains elusive. Herein, garland-like MIL-101(Fe)/$MoS_2$ nanosheets (MMS) with dual metal active sites (Fe and Mo) and rich sulfur vacancies were fabricated to directly activate peroxymonosulfate (PMS) for fast degradation of different organic pollutants (phenols, dyes and drugs), even in real water bodies. The MMS exhibited extremely fast catalytic rate constant of 0.289 $min^{-1}$ in the degradation of sulfamethoxazole (SMX), which was about 36 and 29 times that of single $MoS_2$ (0.008 $min^{-1}$) and MIL-101(Fe) (0.01 $min^{-1}$). Moreover, MMS with good stability and reusability could reach 92% degradation of SMX after 5 cycles. Quenching experiments and electron spin resonance (ESR) tests revealed that hydroxyl radicals (·OH) and singlet oxygen ($^1O_2$) were the dominant reactive oxygen species (ROS) for SMX degradation. The integration of experimental works, characterization techniques and density functional theory (DFT) calculations unraveled that the formation of sulfur vacancies in MMS catalyst could expose more Mo sites, improve the charge density and boost the electron transfer, which was conducive to accelerating the $Fe^{3+}/Fe^{2+}$ cycle for enhancing the activation of PMS. Finally, the C-N, N-O, S-N, C-O and C-S bonds of SMX were easily attacked by ROS to generate the nontoxic intermediates in the MMS/PMS/SMX system. This study offers a new approach to designing high-performance Fe-$MoS_2$ catalysts for the removal of organic pollutants.

## 1. Introduction

With the social development, antibiotics as a member of pharmaceuticals and personal care products (PPCPs) have been widely used to prevent bacterial infections in daily life [1, 2]. Sulfamethoxazole (SMX) is mainly used to prevent urinary tract infections, respiratory infections, and intestinal infections [3]. It has been reported that only a small part of SMX can be absorbed by the human body, while most of the rest is discharged into the water environment as original drugs or major metabolites, which is permanent harm to human health and ecological balance [4]. Therefore, it is urgent to develop an effective method to remove SMX.

Advanced oxidation processes (AOPs) have been widely considered as potential technologies for removing antibiotics from wastewater by generating strong reactive oxygen species (ROS) [5, 6]. Generally, peroxymonosulfate (PMS) as a more popular oxidant is exploited in AOPs than $H_2O_2$ and peroxydisulfate (PDS) due to its advantages: (i) The O-O bond of PMS with asymmetrical structure is more prone to breakage. (ii) PMS as a solid oxidant is safer and easier to store than $H_2O_2$ [7, 8]. Therefore, PMS-based AOPs (PMS-AOPs) is a promising Fenton-like technology owing to its high efficiency, simple operation and low environmental toxicity [9, 10]. At present, it has been reported that transition metal (TM) materials can activate PMS efficiently, which is closely related to the effective filling number of $e_g$ orbitals in 3d electrons of TM ions [11]. Moreover, iron as an earth-abundance, non-toxic and low-cost TM has been used as an excellent activator for PMS activation. Based on the above background, the $Fe^{2+}$/PMS-based AOPs system has a potential application in removing antibiotics due to the faster reaction rate (**Eq. (1-3)**) [12, 13]. However, the $Fe^{2+}$/PMS system still faces some problems such as poor $Fe^{3+}/Fe^{2+}$ cycle, narrow pH range, slow reaction rate, and low utilization of PMS. Especially, the low conversion efficiency of $Fe^{3+}$ and $Fe^{2+}$ can not only limit the removal of organic pollutants in the $Fe^{2+}$/PMS system, but also result in the secondary iron sludge pollution formed by the high concentration of iron leaching [14-16].

$$Fe^{2+} + HSO_5^- \rightarrow SO_4^{\cdot -} + OH^- + Fe^{3+} \qquad k = 3 \times 10^4\ M^{-1}\,s^{-1} \qquad (1)$$

$$Fe^{2+} + H_2O_2 \rightarrow \cdot OH + OH^- + Fe^{3+} \qquad k = 76\ M^{-1}\,s^{-1} \qquad (2)$$

$$Fe^{2+} + S_2O_8^{2-} \rightarrow SO_4^{\cdot -} + SO_4^{2-} + Fe^{3+} \qquad k = 30\ M^{-1}\,s^{-1} \qquad (3)$$

Recently, to accelerate the $Fe^{3+}/Fe^{2+}$ cycle, many attempts have been devoted to introducing co-catalysts into the $Fe^{2+}$/PMS system. Although reducing agents such as homogeneous hydroxylamine [17, 18], citrate [19], cysteine [20] and epigallocatechin gallate [21] are used as co-catalysts to reduce $Fe^{3+}$ to $Fe^{2+}$ for enhancing the degradation efficiency of antibiotics, these reducing agents are easy to cause secondary pollution of water body due to their inherent defects such as toxicity, non-recyclability and high content of total organic carbon. Compared with the homogeneous reductants, the synthetic heterogeneous metal sulfide co-catalysts such as $WS_2$, $CoS_2$, ZnS and $MoS_2$ have drawn greater attention in the PMS-AOPs system, which is attributed to that they can not only be easily recycled after reaction, but also expose more active sites to accelerate the $Fe^{3+}/Fe^{2+}$ cycle [22, 23]. Among these co-catalysts, $MoS_2$ as a typical inorganic metal sulfide co-catalyst, has been widely used in the $Fe^{2+}$/PMS system due to its abundant active sites of surface-exposed $Mo^{4+}$ and electron transfer capability [24, 25]. For example, Wang et al. [26] reported that the enhanced degradation efficiency of organic pollutants in the $MoS_2/Fe^{2+}$/PMS system was due to $MoS_2$ accelerating the transformation of $Fe^{3+}$ to $Fe^{2+}$. Similarly, Xie et al. [27] found the addition of $MoS_2$ was beneficial to the transformation of $Fe^{3+}$ to $Fe^{2+}$ for efficient SMX degradation. Unfortunately, iron ion including $Fe^{2+}$ and $Fe^{3+}$ as a homogeneous phase in the $MoS_2$/PMS system will still face the risk of secondary pollution of iron sludge in the solution.

Inspired by these facts, designing Fe-$MoS_2$ catalysts with high performance for the direct activation of PMS is a promising strategy due to the following strengths: (i) Providing more active sites for PMS activation ($Fe^{3+}$, $Fe^{2+}$, $Mo^{4+}$, $Mo^{6+}$ and sulfur vacancies). (ii) In-situ promoting the $Fe^{3+}/Fe^{2+}$ cycle to reduce metal leaching, achieve higher ion utilization, and form a large amount of ROS. (iii) Separating Fe-$MoS_2$ heterogeneous catalysts from solution conveniently. Recently, Huang et al. [28] reported single Fe atoms doped $MoS_2$ ($Fe_xMo_{1-x}S_2$) as high reactive heterogeneous catalysts for propranolol degradation via sulfite activation due to the synergistic effect of Mo and Fe double catalytic sites of $Fe_xMo_{1-x}S_2$. Li et al. [29] constructed a novel goethite-$MoS_2$ hybrid catalyst with dual active sites to activate PMS for the removal of tetracycline (TC). In addition, to further improve the recovery capacity of Fe-$MoS_2$, Zhou [30] et al. integrated magnetic $Fe_3O_4$ nanoparticles with $MoS_2$ to form a series of magnetic $Fe_3O_4$@$MoS_2$ catalysts, which could remove 100% of sulfonamides within 15 min by PMS activation. Although the Fe-$MoS_2$/PMS system greatly promotes the degradation of antibiotics, it still faces the following challenges: (i) Limited active sites. (ii) The elusive relationship between Fe-$MoS_2$ structure and its catalytic performance. (iii) Lack of antibiotics toxicity evaluation.

In this study, we constructed garland-like MIL-101(Fe)/$MoS_2$ nanosheets (MMS) with dual metal active sites ($Fe^{3+}/Fe^{2+}$ and $Mo^{6+}/Mo^{4+}$) and rich sulfur vacancies via a facile two-step hydrothermal method to directly boost PMS activation for highly efficient degradation of SMX. The unique garland-like structure was composed of MIL-101(Fe) and $MoS_2$ array, which could expose more reaction active sites, improve mass transfer efficiency mass and boost electron transfer. In addition, the introduced sulfur vacancies and Mo sites can accelerate the $Fe^{3+}/Fe^{2+}$ cycle for PMS decomposition to produce the dominant ROS (·OH and $^1O_2$). More specifically, the influence of different MIL-101(Fe) loadings amounts, various factors (e.g., catalyst dosage, PMS dosage, solution pH, and SMX concentration), coexisting anions (e.g., $Cl^-$, $NO_3^-$, $SO_4^{2-}$, $HCO_3^-$ and $H_2PO_4^-$), and organic matters (HA) concentration on SMX degradation kinetics was investigated and optimized. Based on the electrochemical analysis and DFT calculations, the main active sites of MMS for PMS activation were proved, and the underlying role of sulfur vacancies was explained. Finally, the possible degradation pathways of SMX were inferred and the toxicities of intermediates were studied.

## 2. Experimental section

### 2.1. Synthesis of catalysts

### 2.1.1. Synthesis of MIL-101(Fe)

All the chemicals were listed in **Text S1**. The MIL-100(Fe) was synthesized using a previously reported procedure with modifications [31]. In a typical synthesis, 0.675 g of $FeCl_3 \cdot 6H_2O$ was dissolved in 30 mL N,N-dimethylformamide (DMF) to form a homogeneous solution. Next, 0.206 g of terephthalic acid was added to the above solution under stirring for 10 min and sonicating for 10 min. The obtained mixture was transferred into a Teflon liner and kept at 110 °C for 20 h, and cooled naturally. Finally, the yellow product was collected through centrifugation at 10000 rpm and washing three times with DMF and ethanol. The washed material was dried at 60 °C in a vacuum to obtain the MIL-100(Fe) for further use.

### 2.1.2. Synthesis of MIL-101(Fe)/$MoS_2$ catalyst

MIL-101(Fe)/$MoS_2$ (MMS) with diverse loading of MIL-101(Fe) were prepared using a modified two-step hydrothermal method. In brief, a certain amount of MIL-101(Fe) was added into 70 mL ultrapure water and sonicated for 30 min to exfoliate the MIL-101(Fe) nanoparticles for preparing a homogeneous solution. Subsequently, 1.24 g of ammonium molybdate tetrahydrate and 1.06 g of thiourea were added to the above solution under stirring for 30 min. The obtained homogeneous solution was transferred into a Teflon liner and heated at 180 °C for 12 h. After the steel reactor was allowed to cool naturally, the black precipitate was collected through centrifugation and washed several times using deionized water and ethanol. Finally, the washed black solid material was dried in a vacuum oven at 80 °C for 12 h to prepare the MMS-x (where, x represents the mass of MIL-101(Fe) in the composite, x = 0.05, 0.1 and 0.2 g). Pure $MoS_2$ was synthesized via the same method except for the introduction of MIL-101(Fe). Characterization methods of catalysts and experimental procedures were supplied in **Text S2** and **Text S3**, respectively.

### 2.2. Analytical methods

The SMX concentration was measured using a HPLC (Shimadzu LC-16, Japan) (**Fig. S1**) with a UV detector and a CTO-16L column (4.6 mm × 150 mm × 5 μm). The mobile phase was a mixture of acetonitrile and 0.1% formic acid (40:60, V/V) with a flow rate of 1.0 mL/min, and a 10 μL volume of the sample was injected. The detection wavelength and temperature were 272 nm and 30 °C, respectively. The analytical methods of other pollutants were provided in **Table S1**. The intermediates of SMX degradation were investigated by Triple Quad LC/MS (Agilent Technologies, 1290-6460). The residual concentration of PMS in the solution was analyzed using a low-concentration iodide method (**Text S4**). Total organic carbon (TOC) was determined by a TOC analyzer (Vario TN/TOC, elementary, Germany). The relevant electrochemical properties were provided in **Text S5.** The electron spin resonance (ESR, A300-10/12, Bruker) spectroscopy was investigated with DMPO and TEMP as spin-trapping agents to determine ROS (**Text S6**).

## 3. Results and discussion

### 3.1. Catalyst characterizations

The preparation procedures of MIL-101(Fe)/$MoS_2$ catalysts were depicted in **Fig. 1a**. After this, scanning electron microscopy (SEM) was first used to study the morphologies of the synthesized catalysts. MIL-101(Fe) clearly showed an octahedral crystal morphology with a smooth surface. Meanwhile, the mean diameter and standard deviation (SD) of MIL-101(Fe) were 1.55 and 0.33 μm by analyzing SEM images with Image J software (**Fig. S2a**). In addition, $MoS_2$ presented a flower-like structure with a diameter of approximately 1.41 μm, which was formed by stacking and layering thin nanosheets (thickness: 50 nm) (**Fig. S2b**). For the case of MMS-0.1, it was formed through the construction of garland-like nanosheets by arraying MIL-101(Fe) and $MoS_2$. As illustrated in **Fig. S2c**, the synthesized MMS-0.1 roughly retained the $MoS_2$ flower-like structure composed of nanosheet arrays with a rough surface. Moreover, the diameter of the MMS-0.1 became larger compared to MIL-101(Fe) and $MoS_2$, which was attributed to the high dispersion of MIL-101(Fe) in $MoS_2$. Transmission electron microscopy (TEM) was carried out to further study the feature of MMS-0.1. As displayed in **Fig. 1b-d**, the thin square-like structure of MIL-101(Fe) was embedded in the petal-like $MoS_2$, suggesting that two individual components were well combined rather than being independent. Therefore, the unique garland-like structure of MMS-0.1 composed of MIL-101(Fe) and $MoS_2$ array was seen in **Fig. 1c**. The high-resolution TEM (HR-TEM) of MMS-0.1 was shown in **Fig. 1e**. The darker wrinkle morphology with a lattice spacing of 0.62 nm is corresponding to the (002) plane of $MoS_2$. In addition, the lattice spacing with 0.27 nm was ascribed to the (100) face of $MoS_2$ [29, 32]. The selected area electron diffraction (SAED) image of MMS-0.1 was shown in **Fig. 1f.** Two diffraction rings of (002) and (100) could be attributed to $MoS_2$ crystals, which was perfectly consistent with the HR-TEM results. Scanning TEM (STEM) image and the energy dispersive spectroscopy (EDS) mapping of MMS-0.1 were depicted in **Fig. 1g-l**. It showed a uniform distribution of C, Fe, Mo, S, and O in MMS-0.1, further proving the formation of MMS-0.1 nanosheets consisting of MIL-101(Fe) and $MoS_2$.

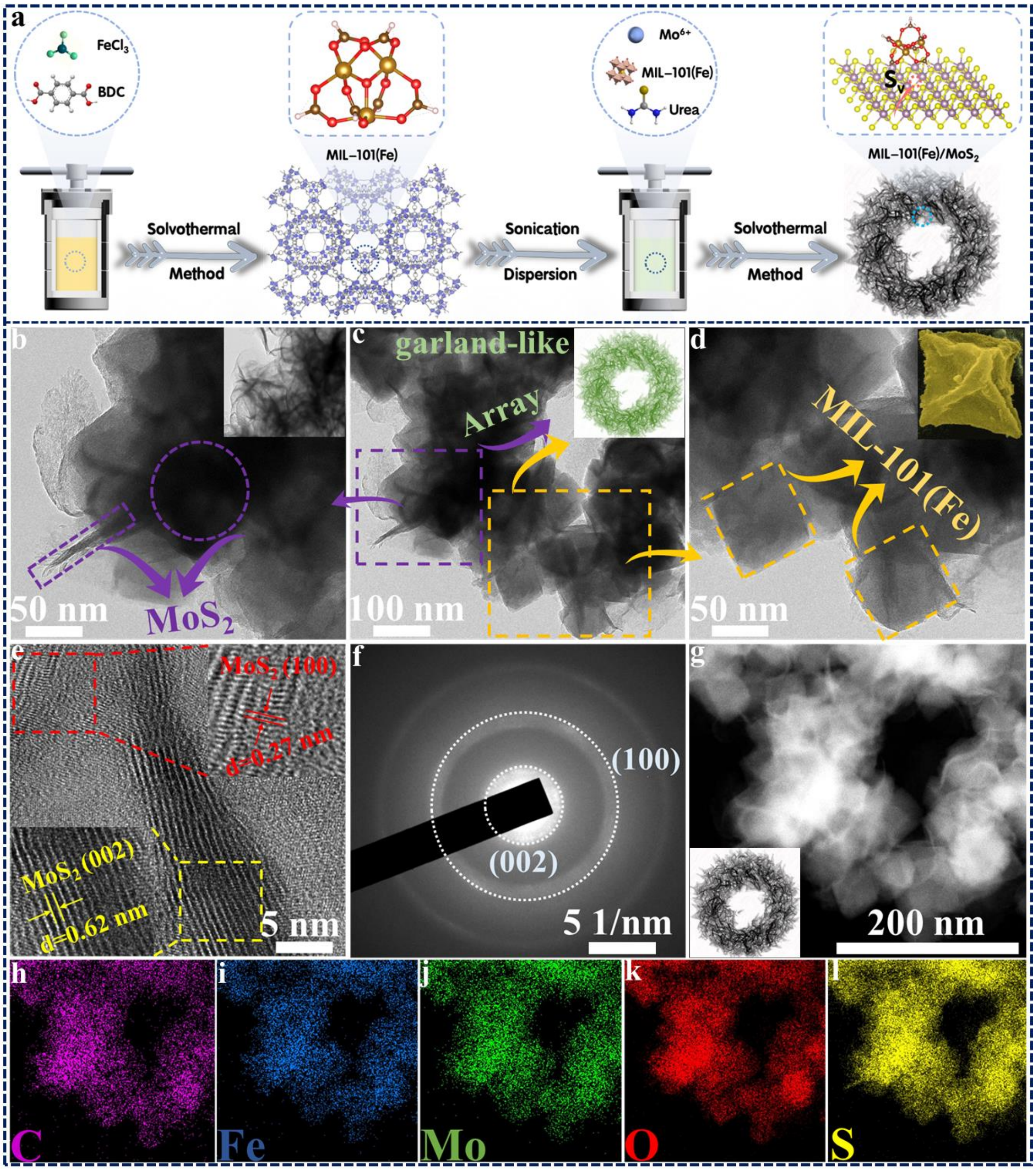

**Fig. 1.** (a) Schematic of preparation of MIL-101(Fe)/$MoS_2$ nanosheets; TEM (b-d), HR-TEM (e), SAED (f), STEM (g) images and corresponding EDS elemental mapping images (h-l) of MMS-0.1.

The physical photographs of MIL-101(Fe) (yellow power with fine particles), $MoS_2$ (black powder with rough particles) and MMS-0.1 (black powder with fine particles) were displayed in **Fig S3**. To explore the crystalline structure of these as-obtained materials, the XRD spectra were revealed in **Fig. 2a**. As for the $MoS_2$, four weak diffraction peaks about at 2θ = 14.4°, 32.7°, 39.5° and 58.3° were attributed to the (002), (100), (103) and (110) crystalline planes (PDF#37-1492), respectively. The diffraction peaks of MIL-101(Fe) were in tune with the simulation MIL-101 pattern with a CCDC number of 605510 [33]. Fortunately, the characteristic diffraction peaks of $MoS_2$ and MIL-101(Fe) were both found in MMS-0.1. However, the peak intensity of $MoS_2$ and MIL-101(Fe) in MMS-0.1 was weak. This might be due to the high dispersion of MIL-101(Fe), which was in agreement with the results of TEM.

To investigate the surface bond structure of the as-obtained catalysts, the Raman spectra of the MIL-101(Fe), $MoS_2$ and MMS-0.1 were performed in **Fig. 2b**. Two peaks were observed at 378 and 408 $cm^{-1}$ for the pure $MoS_2$, which belonged to the out-of-plane mode ($A_g^1$) and in-plane vibration ($E_{2g}^1$), respectively, indicating the existence of 2H $MoS_2$. Additionally, 153.2 $cm^{-1}$ ($J_1$) and 283.6 $cm^{-1}$ ($J_2$) appeared at $MoS_2$ and MMS-0.1, suggesting that 1T $MoS_2$ might be also formed in the MMS-0.1 [34, 35]. To further identify the functional groups and the chemical bonds of as-prepared samples, the fourier transform infrared spectra (FT-IR) were carried out (**Fig. 2c**). The peak at 3380 $cm^{-1}$ was corresponding to the stretching vibrations of the O-H group of the adsorbed $H_2O$ molecules. The adsorption bands at 483, 903, 940, and 1620 $cm^{-1}$ were associated with the stretching vibration of Mo-S, S-S bond, Mo=O band and Mo-O for the $MoS_2$ sample, respectively [36, 37]. The bands of 1599 and 1393 $cm^{-1}$ could be attributed to the asymmetrical and symmetrical stretching of O-C-O in MIL-101(Fe). The peaks at 590 and 790 $cm^{-1}$ were assigned to the vibration of Fe-O and bending vibrations of C-H, respectively, which were the typical characteristic peaks of MIL-101(Fe) [31]. Meanwhile, the characteristic peaks of $MoS_2$ and MIL-101(Fe) were displayed in the MMS-0.1 catalyst, which confirmed the successful synthesis of MIL-101(Fe)/$MoS_2$ nanosheets. In addition, **Fig. S4** showed the FT-IR spectra of MMS-0.05 and MMS-0.2. TGA test at 50-800 °C under the $N_2$ atmosphere (**Fig. S5**) revealed that MMS-0.1 exhibited better thermal stability than MIL-101(Fe) and $MoS_2$, which was due to the synergistic effect of MIL-101(Fe) and $MoS_2$ in MMS-0.1. Furthermore, **Fig. S6a** displayed the $N_2$ adsorption-desorption isotherm of MIL-101(Fe), $MoS_2$ and MMS-0.1. The specific surface area (SSA) and pore volume (PV) of MMS-0.1 was 7.7 $m^2/g$ and 0.028 $cm^3/g$, which exhibited a weaker SSA and PV than those of MIL-101(Fe) (31.88 $m^2/g$ and 0.045 $cm^3/g$) and $MoS_2$ (16.96 $m^2/g$ and 0.042 $cm^3/g$). This phenomenon might be attributed to that anchoring MIL-101(Fe) into $MoS_2$ occupied the porous structure of MMS-0.1. **Fig. S6b** and **Table S2** showed that the pore size of the as-obtained samples appeared at the range of 2-50 nm, suggesting the presence of some mesoporous structures, which was favorable for mass transfer [38].

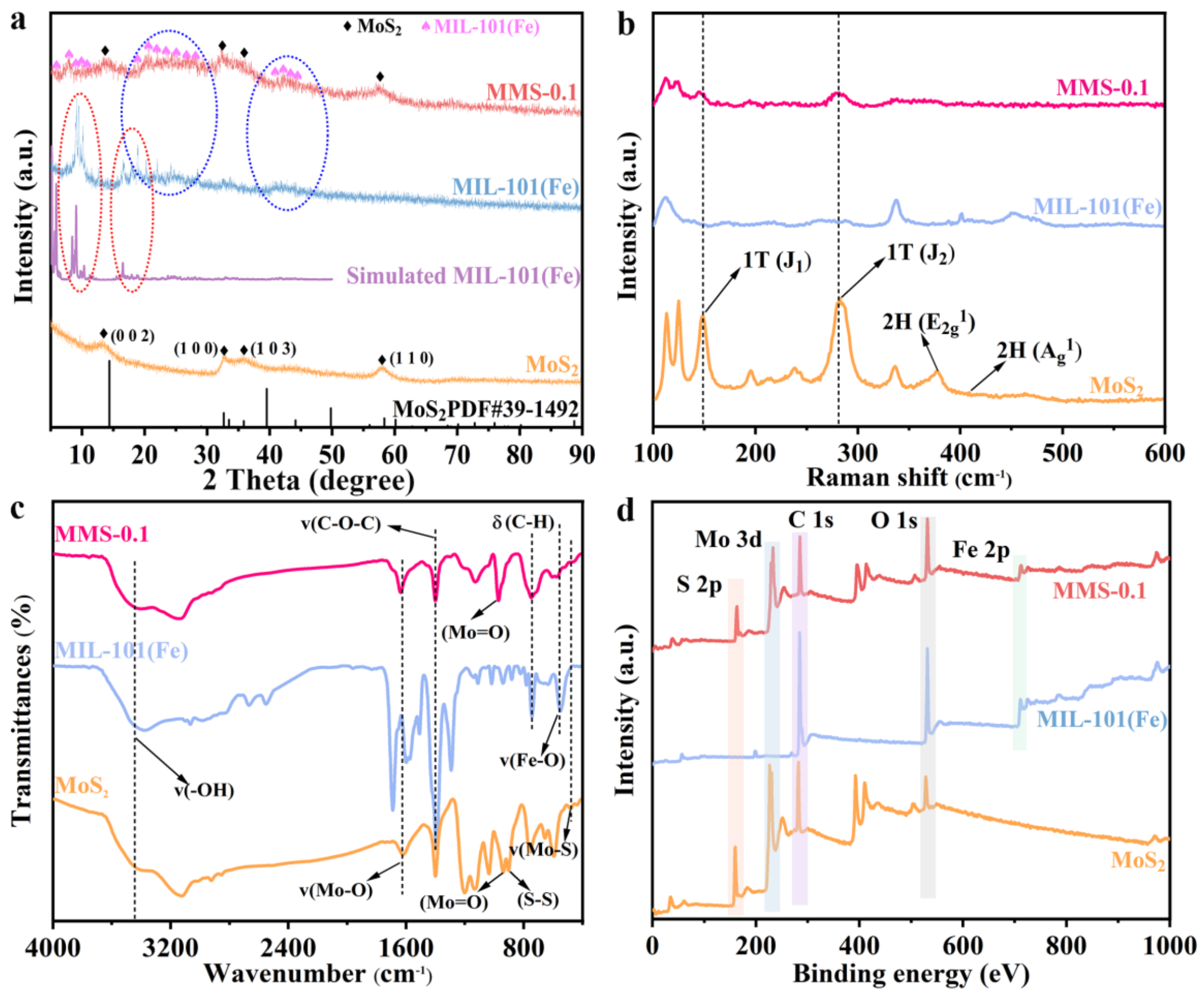

**Fig. 2.** The XRD patterns (a), Raman spectra (b), FT-IR spectra (c), and XPS survey spectra (d) of $MoS_2$, MIL-101(Fe) and MMS-0.1.

The chemical composition and element valence of MIL-101(Fe), $MoS_2$ and MMS-0.1 were studied using X-ray photoelectron spectra (XPS). **Fig. 2d** revealed the existence of five elements including C, O, S, Mo and Fe in MMS-0.1, suggesting the successful recombination of MIL-101(Fe) with $MoS_2$. Moreover, **Fig. 3a** displayed the atomic ratio of, S, Mo and Fe of all catalysts. **Table S3** showed that the atomic ratio of C, O, S, Mo and Fe in MMS-0.1 was 53.29, 25.77, 10.28, 8.44 and 2.22 at%, respectively. To further explore the chemical state, the Fe 2p, S 2p and Mo 3d spectra of catalysts were analyzed. **Fig. 3b** showed the high-resolution Fe 2p spectrum. The peaks at 710.57 and 724.17 eV were corresponding to the $2p_{3/2}$ and $2p_{1/2}$ of $Fe^{2+}$, and the peaks at 712.96 and 726.85 eV were attributed to the $2p_{3/2}$ and $2p_{1/2}$ of $Fe^{3+}$, suggesting the presence of $Fe^{2+}/Fe^{3+}$ in MMS-0.1 [29]. In addition, the satellite peaks were found at 720.05 and 733.01 eV. For the S 2p spectrum in **Fig. 3c**, the binding energy of S $2p_{3/2}$ (161.59 eV) and S $2p_{1/2}$ (162.7 eV) could be assigned to the orbitals of $S^{2-}$, suggesting the sulfur species and Fe-S bonds were probably present at the surface of MMS-0.1 [30]. Besides, the peaks located at 168.83 and 163.36 eV could be identified as sulfate species ($-SO_n-$) and unsaturated edge S, respectively [36]. The high-resolution Mo 3d spectra of MMS-0.1 could be deconvoluted into four peaks at 226.06, 229.00, 232.47, and 236.00 eV, respectively [39]. The main peaks at 229.00 and 232.47 eV corresponded to Mo $3d_{5/2}$ and Mo $3d_{3/2}$, suggesting the presence of $Mo^{4+}$ in MMS-0.1 (**Fig. 3d**). The small peak at 225.71 eV was attributed to S 2s because of the Mo-S bonding. The other peak at 336 eV could be assigned to the $Mo^{6+}$ oxide species formed during the in-situ synthesis of MMS-0.1. **Fig. S7** revealed that the high-resolution of O 1s XPS spectrum at 531.57, 532.39 and 533.13 eV belonged to Mo-O, Fe-O and -COOH, respectively, indicating the preservation of $MoS_2$ and MIL-101(Fe) in MMS-0.1 [32]. The C 1s spectrum was shown in **Fig. S8**, and the binding energy of 284.67 eV was attributed to the carboxyl group (O=C-O) originating from the carboxyl species of the terephthalic acid ligand. Based on the abovementioned results, multiple active sites such as $Fe^{3+}/Fe^{2+}$, $Mo^{6+}/Mo^{4+}$ and sulfur species coexisted in MMS-0.1.

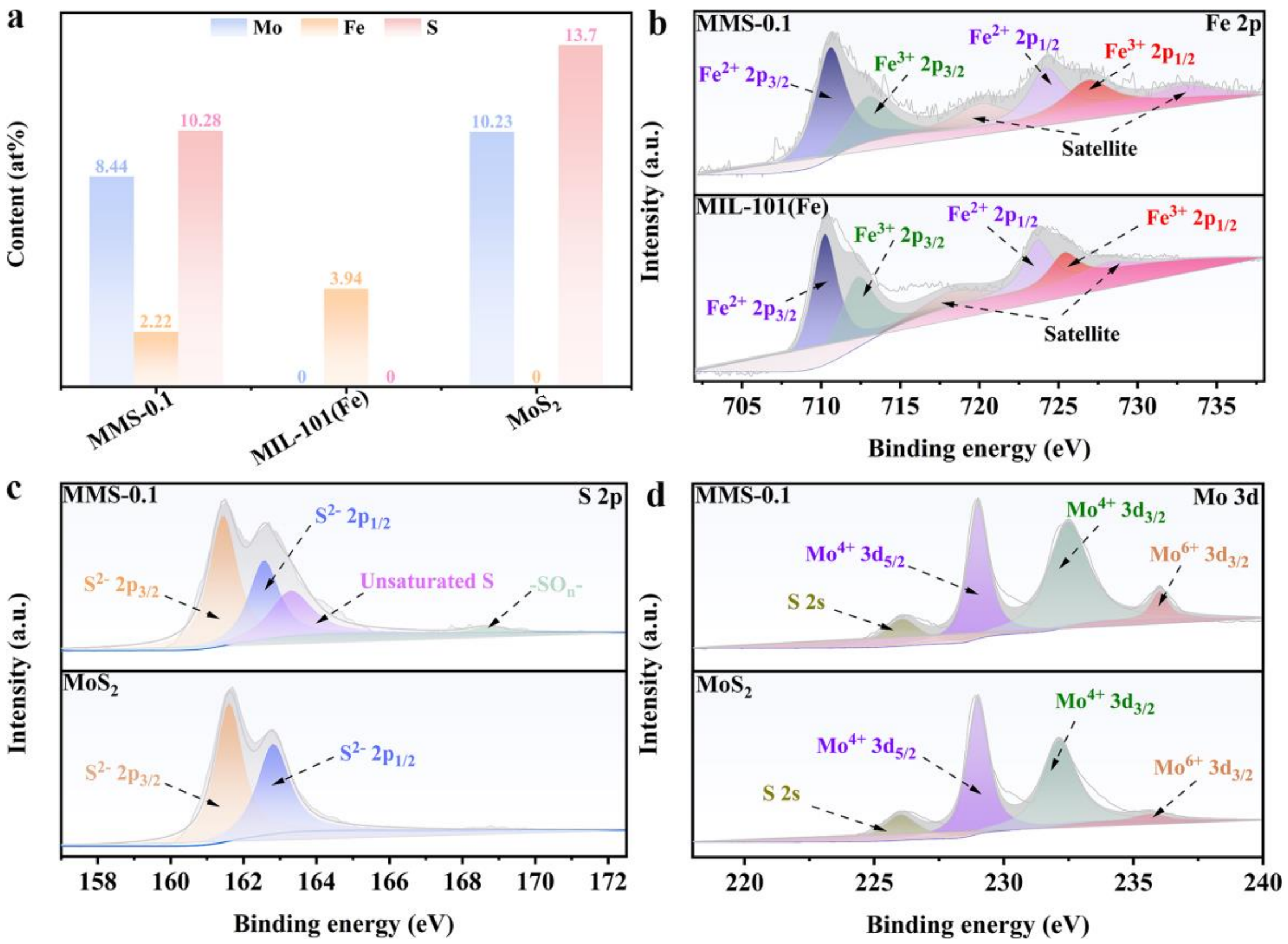

**Fig. 3.** The surface elemental contents of catalysts (a), high-resolution Fe 2p spectra (b), S 2p spectra (c), Mo 3d spectra (d).

## 3.2. Catalytic performance of MMS

Before evaluating the catalytic degradation of SMX using the as-obtained catalysts, the adsorption capacity was first investigated. As illustrated in **Fig. S9**, the limited adsorption efficiency of SMX was found at the range of 6.3%-9.6% within 30 min by all catalysts, suggesting that the adsorption effect of catalysts was negligible in the process of SMX removal. Furthermore, **Fig. 4a** evaluated the catalytic performance of these as-prepared catalysts by activating PMS for SMX degradation. It could be observed that about 20% of SMX was eliminated in the presence of PMS due to its limited self-activation, indicating that PMS was difficult to be activated without external assistance. Moreover, 27%, 35%, 95%, 100% and 100% of SMX were degraded within 30 min in the $MoS_2$/PMS, MIL-101(Fe)/PMS, MMS-0.05/PMS MMS-0.1/PMS and MMS-0.2/PMS, respectively. The results displayed that MMS-0.1 as a heterogeneous catalyst exhibited high performance for the degradation of SMX, which was superior to previously reported metal-based composites (**Table S4**). Meanwhile, the pseudo-first-order kinetics of different systems were fitted in **Fig. S10a**. The corresponding SMX degradation rate constants were shown in **Fig. 4b**. The MMS-0.1/PMS obtained the rate constant of 0.289 $min^{-1}$, which was 36.1 and 28.9 times higher than those of $MoS_2$/PMS and MIL-101(Fe)/PMS, respectively. In addition, the PMS consumption rate in different systems was measured in **Fig. 4c**. Over 90% of PMS was decomposed in MMS-0.1/PMS and MMS-0.2/PMS systems, further proving the excellent catalytic activity of MMS-0.1 and MMS-0.2. Although the *k* of MMS-0.2 was similar to that of MMS-0.1, less MIL-101(Fe) was consumed during the preparation of MMS-0.1. Therefore, considering the concept of green chemistry, MMS-0.1 was selected as the typical catalyst to continue the following experiments.

To gain insights into the mineralizing performance of MMS-0.1, the TOC removal rate was studied in the MMS-0.1/PMS/SMX system as the reaction time increased (**Fig. 4d**). The TOC removal rate of 46.2% was obtained at 30 min, suggesting that the higher mineralizing rate might be dependent on the radical pathway rather than the non-radical pathway. Meanwhile, about 142 mg/L of $SO_4^{2-}$ was detected after reaction via an ion chromatography, which was lower than the limited standard of $SO_4^{2-}$ concentration ( < 250 mg/L) in potable water of China [40].

Subsequently, the leaching concentration of metal ions increased as the reaction time increased (**Fig. 4e**). The leaching concentration of Fe and Mo ions was 1.02 and 2.62 mg/L in 30 min, respectively, accounting for 1.02% and 2.62% of the MMS-0.1 dosage (0.1 g/L), which was lower than that of previous reports [26, 39]. **Fig. 4f** showed that 99%, 91% and 79% of SMX were degraded by PMS, PS and $H_2O_2$ activation within 30 min, respectively. The $k$ values followed the order of PMS ($k$ = 0.289 $min^{-1}$) > PS ($k$ = 0.111 $min^{-1}$) > $H_2O_2$ ($k$ = 0.056 $min^{-1}$) (**Fig. S10b**). This result determined the optimum peroxide for PMS in the MMS-0.1/SMX system because the O-O bond of PMS was easily broken due to its asymmetric structure [41]. Furthermore, **Fig. 4g** and **Fig. S10c** investigated that tetracycline hydrochloride (TC), bisphenol A (BPA), rhodamine B (RhB) and sulfadiazine (SDZ) were also quickly degraded in the MMS-0.1/PMS system, exhibiting the universal applicability of MMS-0.1. **Table S5** summarized the corresponding properties of various organic pollutants. After five successive cycles, the degradation rate of SMX still reached 92 % in 30 min without extra regeneration treatment, and there was no significant change in the XRD spectra of MMS-0.1, suggesting the good stability and reusability of MMS-0.1 (**Fig. 4h and Fig. S11**). The corresponding $k$ values of 1st, 2nd, 3rd, 4th and 5th were 0.29, 0.19, 0.148, 0.128 and 0.08 $min^{-1}$, respectively (**Fig. S10d)**. This reduced $k$ value might be due to the coverage of active sites by the intermediates of SMX under the degradation processes, hindering the $Fe^{3+}/Fe^{2+}$ cycle [42]. Meanwhile, the Fe and Mo ions concentration of each cycle after the reaction was also detected in **Fig. S12.** The leached Fe ion concentration of every cycle was lower than the European Union standard of 2 mg/L [43]. For Mo ion, the leached concentration of every cycle was low than 2.8% of the MMS-0.1 dosage (0.1 g/L). This result further suggested the stable chemical structure of the MMS-0.1 catalyst. Finally, in comparison to SMX degradation catalyzed by previously reported metal-based catalysts, it was found that MMS-0.1 had a better potential application in practical wastewater treatment (**Fig. 4i**).

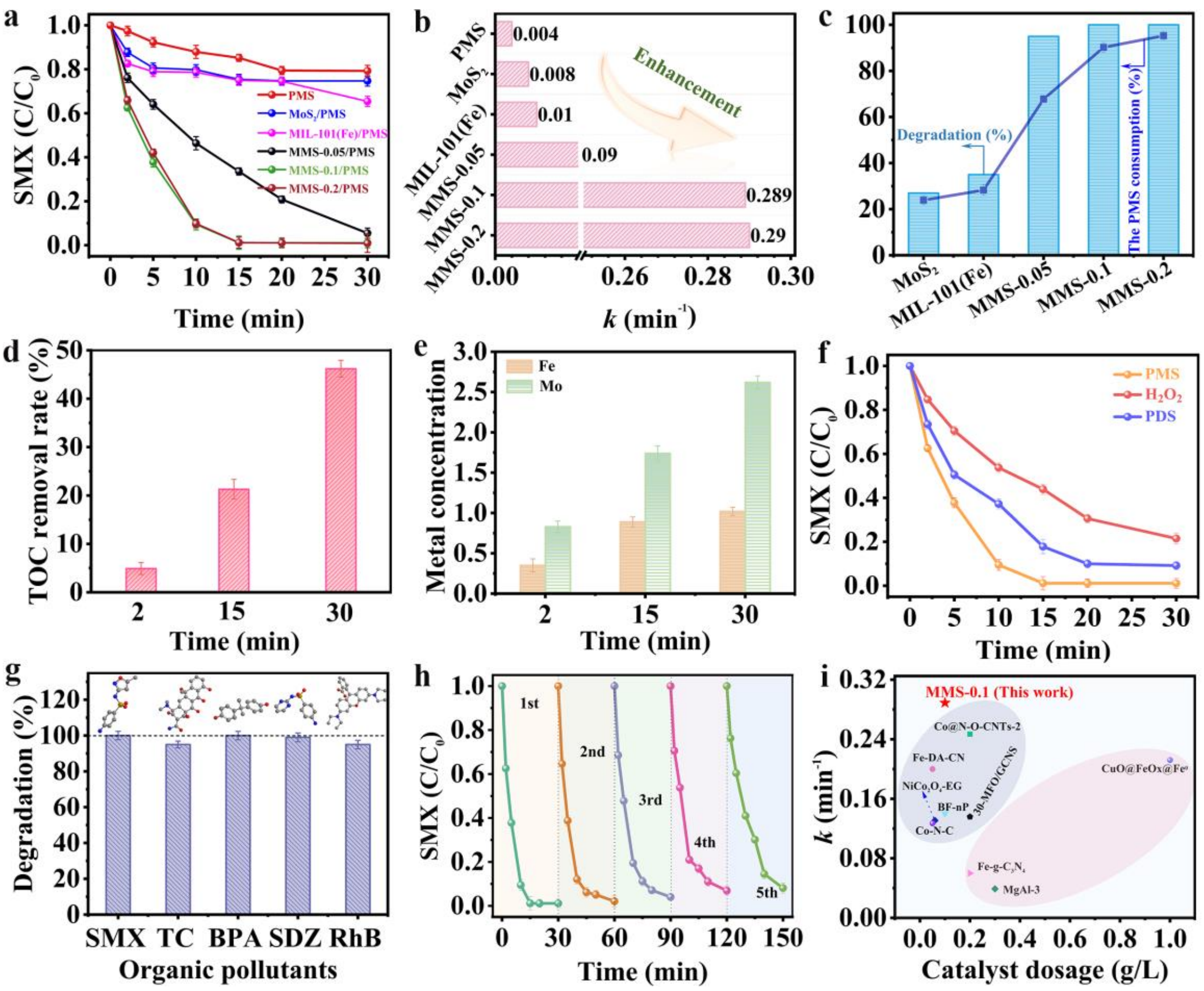

**Fig. 4.** (a) Degradation of SMX by different catalysts; (b) The corresponding SMX degradation rate constants; (c) The PMS consumption rate in different systems; (d) The removal rate of TOC in the MMS-0.1/PMS/SMX system; (e) The concentration of leached metals of MMS-0.1; (f) Various oxidants activation by MMS-0.1; (g) Degradation of various pollutants by the MMS-0.1/PMS system; (h) Evaluation of the reusability of MMS-0.1; (i) Comparison with previous reported metal-based composites. Reaction conditions: [catalyst] = 0.1 g/L, [PMS] = 1.0 mM, initial pH = 6.18, [SMX] = 10 mg/L, T = 25 ± 2 °C.

**3.3. Effects of various factors on SMX degradation**

To further evaluate the catalytic performance of, the various influential factors including catalyst dosage, PMS dosage, initial pH values, and SMX concentration were systematically studied. As shown in **Fig. S13a**, when the MMS-0.1 dosage increased from 0.05 to 0.2 g/L, the degradation efficiency of SMX was gradually improved from 49% to 100% within 10 min, and the corresponding *k* increased from 0.067 to 0.516 $min^{-1}$ (**Fig. 5a** and **Fig. S14a**). Besides, when the MMS-0.1 dosage increased from 0.1 to 0.2 g/L, there was no obvious improvement in the degradation efficiency of SMX in 30 min. However, the corresponding *k* values significantly increased from 0.116 to 0.516 $min^{-1}$, implying that a high MMS-0.1 dosage could provide sufficient active sites for accelerating the degradation of SMX. **Fig. S13b** demonstrated the effects of PMS dosage on SMX degradation. Only 82% of SMX was degraded when the dosage of PMS was 0.25 mM due to PMS depletion. Moreover, when the dosage of PMS increased from 0.5 to 1 mM, the degradation efficiency of SMX increased from 84% to 99% within 15 min and the corresponding *k* value improved from 0.116 to 0.29 $min^{-1}$, respectively. However, when the concentration of PMS further reached 2 mM, the degradation efficiency of SMX and the *k* value both slightly decreased, which could be attributed to the self-scavenging of excessive radical or ineffective PMS decomposition (**Fig. 5a** and **Fig. S14b**).

Therefore, 0.5 mM PMS and 0.1g/L MMS-0.1 were selected as the optimal conditions to evaluate the following experiments.

Generally, pH played an essential role in the degradation of SMX in SR-AOPs. As seen in **Fig. S13c**, 100% of SMX was degraded at pH=3.06 in 10 min, and the corresponding *k* value was 0.376 $min^{-1}$, which was higher than that of the control experiments. Moreover, when the pH value increased from 5.39 to 9.10, the degradation efficiency of SMX and the corresponding *k* hardly changed (**Fig. 5a** and **Fig. S14c**). However, alkaline conditions (pH = 11.45) significantly inhibited SMX degradation due to the rapid self-decomposition rate of PMS [44]. **Fig. S13d** showed the influence of initial SMX concentration, the SMX degradation efficiency was 99%, 84%, 64% and 56% at initial SMX concentration of 5, 10, 15 and 20 mg/L in 15 min, respectively. The corresponding *k* values were 0.178, 0.166, 0.06, and 0.045 $min^{-1}$ (**Fig. 5a** and **Fig. S14d**). It could be seen that the degradation efficiency of SMX was negatively correlated with the initial SMX concentration because the excessive SMX concentration needed to consume a larger ROS.

### 3.4. The anti-interference ability of the MMS-0.1/PMS system

To explore the potential practical applications of MMS-0.1 in wastewater treatment, the effects of typical anions ($Cl^-$, $NO_3^-$, $SO_4^{2-}$, $HCO_3^-$ and $H_2PO_4^-$) and HA on SMX degradation were studied. As shown in **Fig. S15a** and **Fig. 5b**, when the concentration of $Cl^-$ increased from 0 to 5 mM, the degradation efficiency of SMX increased from 85% to 100% in 15 min. Compared with the control experiment ($k$ = 0.116 $min^{-1}$), the *k* values of $Cl^-$ (2 mM) ($k$ = 0.199 $min^{-1}$) had a positive effect on SMX degradation (**Fig. S16a**). This result might be attributed to the fact that $Cl^{-}$ could react with PMS to form other oxidizing species (**Eqs. 4-6**) [45]. However, 10 mM of $Cl^{-}$ showed a similar degradation efficiency (84%) and *k* (1.21min$^{-1}$) in 15 min to control experiment, because excess $Cl^{-}$ could consume $SO_4^{\cdot-}$ and ·OH to produce chloride radicals (·Cl/·$Cl_2^{-}$/·$ClOH^{-}$) with lower oxidation potential followed the (**Eq. 7-10**) [46]. According to **Fig. S15(b-c)** and **Fig. 5b**, the addition of $NO_3^{-}$ and $SO_4^{2-}$ (2-10 mM) had little effect on the degradation of SMX (**Fig. S16b-c**). However, 2-10 mM of $HCO_3^{-}$ had a severe inhibitory effect on the degradation efficiency of SMX. Only about 25% SMX could be degraded and the corresponding *k* value decreased to 0.005 min$^{-1}$ (**Fig. S16d**). The significant inhibitory effect of $HCO_3^{-}$ could be attributed to that: (1) $HCO_3^{-}$ could increase the pH value of the reaction system, which was consistent with the results of **Fig. S13c**. (2) It was reported that $HCO_3^{-}$ could be used as a scavenger of ROS, so $HCO_3^{-}$ might react with the $SO_4^{\cdot-}$ and ·OH to produce the lower oxidation potential of ·$CO_3^{-}$ in the system (**Eq. 11-12**) [47]. According to **Fig. S15e** and **Fig. 5b**, the presence of 2 mM $H_2PO_4^{-}$ inhibited the degradation to 48%, and the corresponding *k* value decreased to 0.018 min$^{-1}$. As the concentration of $H_2PO_4^{-}$ increased from 2 to 10 mM, the degradation date was further reduced to 24%. Meanwhile, the *k* value was 0.007 min$^{-1}$ at 20 mM of $H_2PO_4^{-}$ (**Fig. S16e**). This might be due to that $H_2PO_4^{-}$ could occupy the active sites on the catalyst to prevent contact between PMS and the catalyst, reducing the SMX degradation significantly [48]. Generally, HA as a type of natural organic matter (NOMs) was widely present in natural water. It was well-known that HA could consume the ROS and easily adsorb on the active sites of the catalyst surface, hindering the degradation performance of the catalyst [49, 50]. However, **Fig. S15f** and **Fig. S16f** showed that HA exhibited a negligible effect on SMX degradation in the MMS-0.1/PMS/SMX system due to the weak reactivity of HA with the non-radical pathway species $^1O_2$ at short reaction times [51]. Furthermore, to explore the potential application of the MMS-0.1 in real water bodies, lake water (LW) (taken from the central lake of Guangzhou University City), river water (RW) (taken from the east campus of Sun Yat-sen University), tap water (TW) (taken from our laboratory), and ultrapure water (UP) were used as diverse the medium of SMX (**Table S6)**. As shown in **Fig. S17**, the degradation efficiency of SMX in TP was very close to that of UP, suggesting a negligible negative effect. The degradation efficiency of SMX in RW and LW slightly decreased due to the presence of more organic matter and background ions compared with others. Importantly, over 81% of SMX was degraded in the above water medium, indicating that the MMS-0.1/PMS system was a promising technology for actual antibiotic remediation in the water.

$$Cl^- + HSO_5^- \rightarrow HClO + SO_4^{2-} \quad (4)$$

$$Cl^- + HSO_5^- + H^+ \rightarrow Cl_2 + SO_4^{2-} + H_2O \quad (5)$$

$$Cl^- + SO_4^{\cdot -} \rightarrow \cdot Cl + SO_4^{2-} \quad (6)$$

$$Cl^- + \cdot OH \rightarrow \cdot Cl + OH^- \quad (7)$$

$$Cl^- + \cdot Cl \rightarrow \cdot Cl_2^- \quad (8)$$

$$2\ \cdot Cl_2^- \rightarrow 2\ Cl^- + Cl_2 \quad (9)$$

$$\cdot Cl + OH^- \rightarrow \cdot ClOH^- \quad (10)$$

$HCO_3^{-} + \cdot OH + \rightarrow \cdot CO_3^{-} + H_2O$ (11)

$HCO_3^{-} + SO_4^{\cdot -} \rightarrow \cdot CO_3^{-} + SO_4^{2-} + H^{+}$ （12）

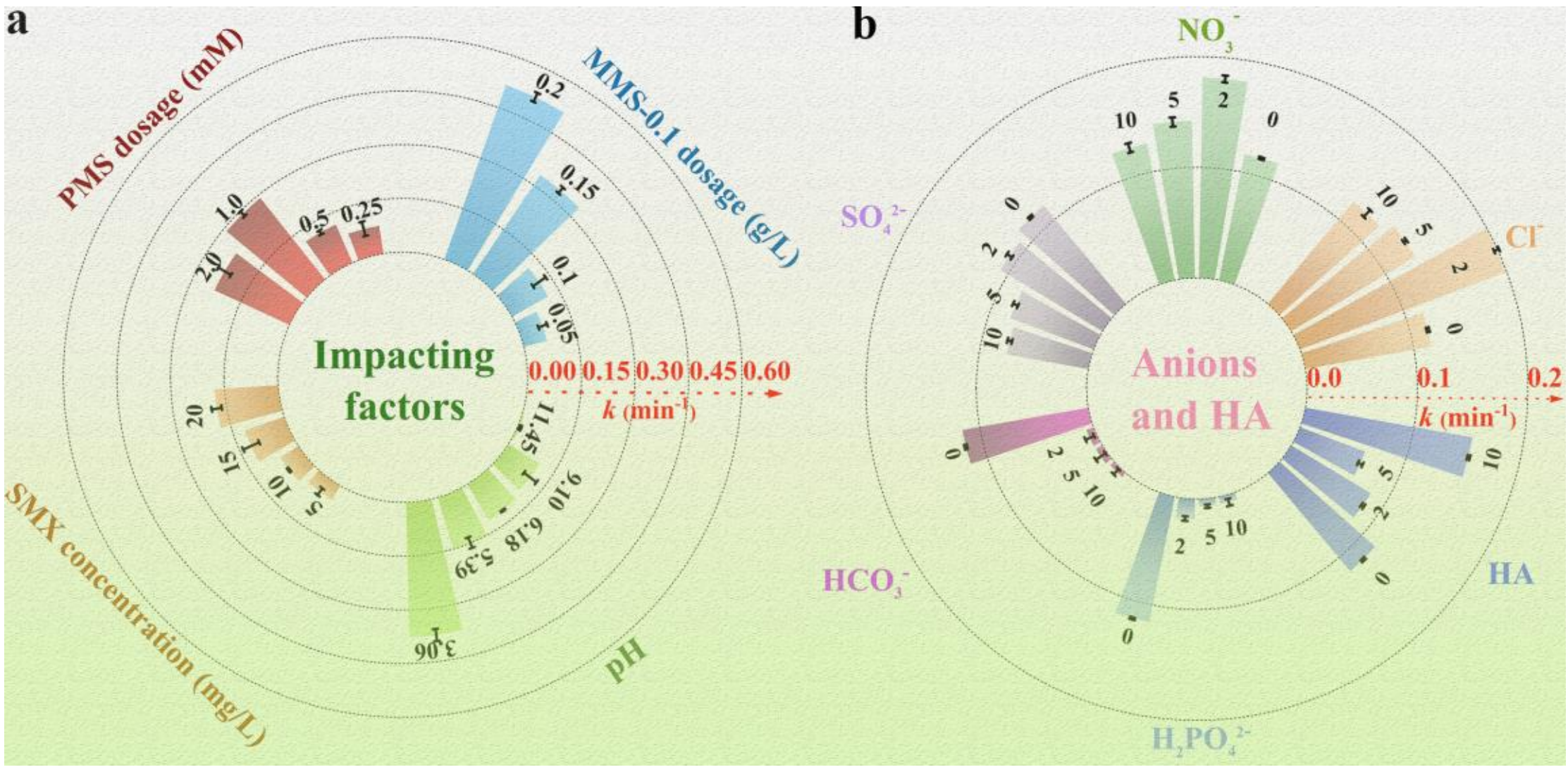

**Fig. 5.** The corresponding rate constants of the effect of impacting factors (a) and anions and HA (b) on SMX degradation. Reaction conditions: [MMS-0.1] = 0.1 g/L, [PMS] = 0.5 mM, initial pH = 6.18, [SMX] = 10 mg/L, T = 25 ± 2 °C.

### 3.5. Reaction mechanism

#### 3.5.1. Identification of reactive oxidative species

To explore the ROS in the MMS-0.1/PMS/SMX system, a series of quenching experiments were carried out. In general, methanol (MeOH) and tert-butyl alcohol (TBA) were chosen as scavengers to confirm the existence of ·OH and $SO_4^{\cdot -}$. MeOH was often used to quench both ·OH ($k$ =1.2-2.8×10$^{8}$ M$^{-1}$ s$^{-1}$) and $SO_4^{\cdot -}$ ($k$ =1.6-7.8×10$^{7}$ M$^{-1}$ s$^{-1}$) simultaneously. While TBA could be used as a specific scavenger for ·OH ($k$ = 3.8-7.6×10$^{8}$ M$^{-1}$ s$^{-1}$) rather than $SO_4^{\cdot -}$ ($k$ = 4-9.1 × 10$^{5}$ M$^{-1}$ s$^{-1}$) [52, 53]. As shown in **Fig. S18**, with the increase of MeOH concentration, the degradation efficiency of SMX was significantly inhibited. When 500 mM MeOH was added to the MMS-0.1/PMS/SMX system, only 32% of SMX was removed in 30 min, and the $k$ decreased from 0.116 to 0.037 $min^{-1}$. Moreover, the high concentration of TBA (500 mM) had a certain inhibitory effect on SMX degradation **Fig. S19**. This result proved that both ·OH and $SO_4^{\cdot-}$ were produced in the system actually, but ·OH made a greater contribution to SMX degradation than $SO_4^{\cdot-}$. Besides, the detection of fluorescence spectra further proved the generation of a large number of ·OH in the system (the detailed discussion in **Text S7** and **Fig. S20**). Benzoic acid (BA) was chosen as a probe molecule to obtain the quantification of cumulative ·OH concentration in the system (**Text S8**). **Fig. S21** displayed that as the reaction time increased, the concentration of ·OH generated in the MMS-0.1/PMS system tended to stabilize at 125.8 μM. After that, p-BQ ($k$ = 9.6 × $10^8$ $M^{-1}$ $s^{-1}$) was chosen as scavengers for $O_2^{\cdot-}$ [25]. As shown in **Fig. 6a** and **Fig. S22**, 91% of SMX could be degraded with the addition of p-BQ (10 mM), implying a rather limited contribution of $O_2^{\cdot-}$ for SMX degradation. Subsequently, furfuryl alcohol (FFA) could react with $^1O_2$ ($k$ =1.2×$10^8$ $M^{-1}$ $s^{-1}$) [54]. It was found that the presence of 10 mM FFA could strongly inhibit the degradation of SMX. The degradation efficiency of SMX was only 12% and the $k$ decreased to 0.0037 $min^{-1}$ (**Fig. 6b**). To further confirm the generation of $^1O_2$ in the MMS-0.1/PMS/SMX system, β-carotene with higher reactivity with $^1O_2$ ($k$ =2-3×$10^{10}$ $M^{-1}$ $s^{-1}$) was chosen as a quencher for $^1O_2$ [55, 56]. As shown in **Fig. S23**, the addition of β-carotene (0.1 mM) resulted in the apparent decrease of both SMX degradation (from 100 to 61%) and $k$ (from 0.116 to 0.034 $min^{-1}$) for the MMS-0.1/PMS/SMX system, proving that $^1O_2$ was also the main contributor to SMX degradation. The steady-state concentration of $^1O_2$ was calculated as 1.90 ×10 $^{-5}$ μM in the MMS-0.1/PMS system (**Fig. S24**).

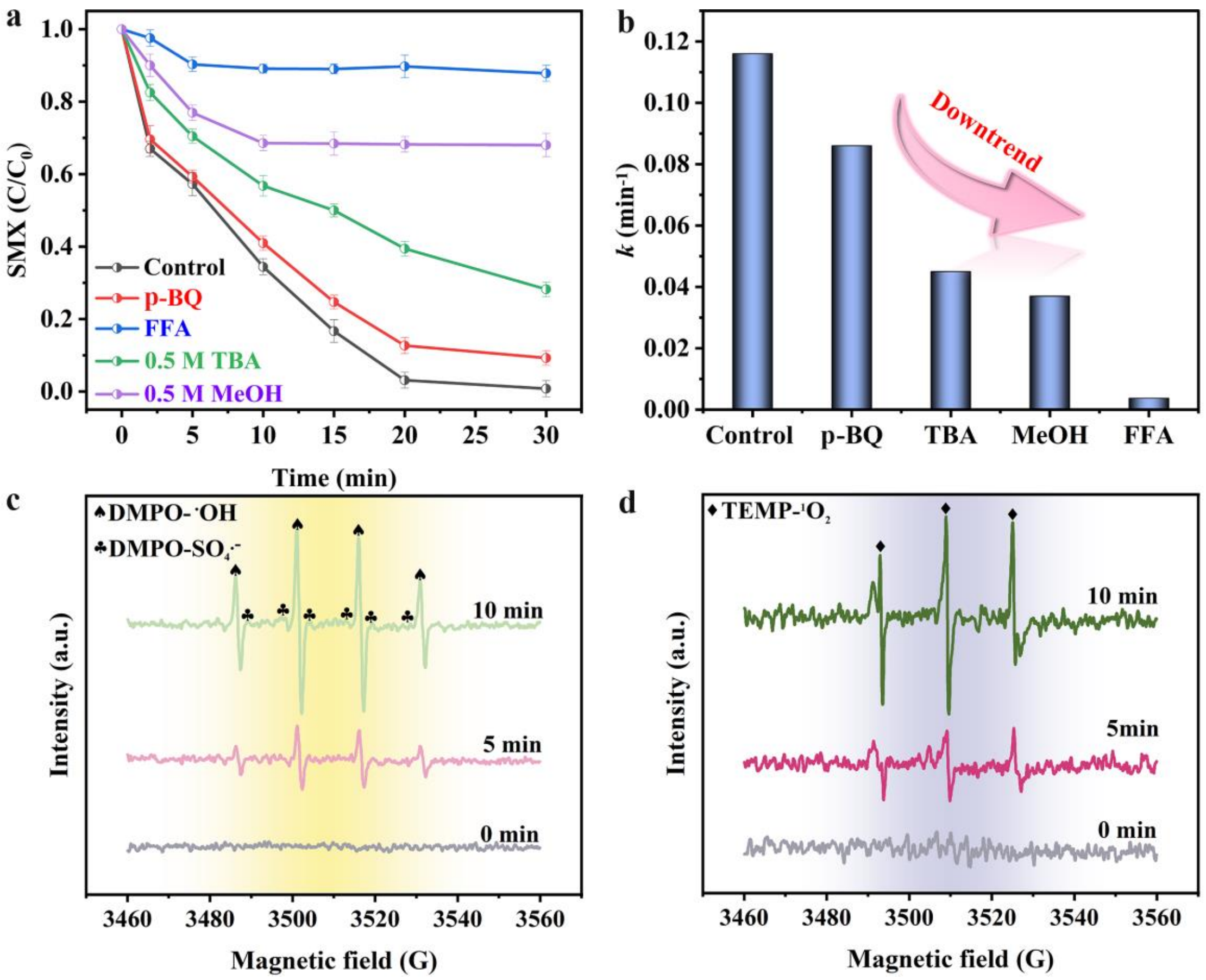

**Fig. 6.** Influence of different scavengers on SMX degradation: (a) MeOH, TBA, FFA and p-BQ. (b) the corresponding SMX degradation rate constants. (c) DMPO-·OH and DMPO-$SO_4^{\cdot-}$, and (d) TEMP-$^1O_2$ at different reaction times in the MMS-0.1/PMS system. Reaction conditions: [MMS-0.1] = 0.1 g/L, [PMS] = 0.5 mM, initial pH = 6.18, [SMX] = 10 mg/L, [FFA] = [p-BQ] = 10 mM, T = 25 ± 2 °C.

The ESR spectroscopy could be used to further investigate the existence of ROS in the MMS-0.1/PMS/SMX system. It was well-known that DMPO was a spin-trapping agent for ·OH and $SO_4^{\cdot-}$, and TEMP could react with $^1O_2$ to produce TEMP-$^1O_2$ [57-59]. As shown in **Fig. 6c**, no DMPO-·OH and DMPO-$SO_4^{\cdot-}$ signals were found without the addition of MMS-0.1 (0 min). Simultaneously, when MMS-0.1 was added to PMS/DMPO solution, the strong characteristic signals of DMPO-·OH with the intensity ratio of 1:2:2:1 were detected with the reaction time. However, the ESR signals of DMPO-$SO_4^{\cdot-}$ were weak, which further proved that ·OH played the dominant role in SMX degradation rather than $SO_4^{\cdot-}$. **Fig. 6d** showed that the signal intensity of a typical triplet signal with 1:1:1 (TEMP-$^1O_2$) increased gradually, suggesting the generation of $^1O_2$ in the MMS-0.1/PMS system. Based on the quenching experiments and ESR test, it could be concluded that ·OH and $^1O_2$ were the dominant ROS for SMX degradation.

### 3.5.2. Activation mechanism of PMS by MMS-0.1

To determine the active sites of MMS-0.1 for producing ROS, XPS spectra of fresh and used MMS-0.1 were carried out. **Fig. 7a** showed the XPS survey spectra of the fresh and used MMS-0.1. The peak intensity of S 2p, Mo 3d and Fe 2p of MMS-0.1 all decreased after the reaction, suggesting that the different valence states of these elements might be the active sites for PMS activation. Meanwhile, the relative content of S 2p, Mo 3d and Fe 2p decreased from 10.28 to 8.37 at%, 8.44 to 7.42 at% and 2.22 to 1.27 at% for the used MMS-0.1 (**Table S7**). As shown in **Fig. 7b**, the four peaks were clear in the Mo 3d spectrum, but the relative content of $Mo^{4+}$ decreased from 82.87% to 75.49% and the content of $Mo^{6+}$ increased from 5.71% to 11.88% after the catalytic reaction, suggesting that $Mo^{4+}$ could react with PMS to produce $Mo^{6+}$ (**Table S7**). However, the decreasing content of $Mo^{4+}$ (7.38%) was different from the increasing content of $Mo^{6+}$ (6.17%), indicating the existence of $Mo^{4+}/Mo^{6+}$ redox interconversions during the degradation experiment. Besides, the Fe 2p spectra were shown in **Fig. 7c**. After the reaction, the Fe 2p envelope of MMS-0.1 became weaker. The percentage of $Fe^{2+}$ dropped by 4.35% with the increase of the percentage of $Fe^{3+}$ by 5.64%, suggesting that the $Fe^{2+}/Fe^{3+}$ cycle was present in the MMS-0.1/PMS system (**Table S7**). In **Fig. 7d** and **Table S7,** the percentage of $S^{2-}$ $2p_{3/2}$ and $S^{2-}$ $2p_{1/2}$ declined from 45.16% and 29.31% to 42.12% and 28.31%, respectively, while the unsaturated S also dropped from 66.1% to 62.6%. However, the content of $-SO_n-$ significantly increased from 5.13% to 12.93%. This result could be deduced that low-valent S played a role in promoting $Fe^{2+}/Fe^{3+}$ and $Mo^{4+}/Mo^{6+}$ cycles [60, 61]. Therefore, the generation process of ROS was as follows: (i) the redox pairs of $Fe^{2+}/Fe^{3+}$ and $Mo^{4+}/Mo^{6+}$ could activate PMS to produce the radicals ($\cdot OH$, $SO_4^{\cdot -}$ and $SO_5^{\cdot -}$) (**Eqs. (13-17)**). In addition, the formed $SO_4^{\cdot -}$ could be further consumed by $H_2O$ for transformation to $\cdot OH$ (**Eqs. (18)**). (ii) Meanwhile, $Mo^{4+}$ could donate electrons to $Fe^{3+}$ for enhancing the regeneration of $Fe^{2+}$ and $Mo^{6+}$ (**Eqs. (19)**). (iii) Furthermore, $Fe^{3+}/Mo^{6+}$ could receive electrons from the low-valent $S^{2-}$ to produce $Fe^{2+}/Mo^{4+}$, respectively (**Eqs. (20)**). (iv) Finally, a large number of $SO_5^{\cdot -}$ could be reacted with $H_2O$ to the generation of $^1O_2$ (**Eqs. (21)**). As a result, the radical ($\cdot OH$) and non-radical ($^1O_2$) pathways played a major contribution to the SMX degradation in the MMS-0.1/PMS system (**Eqs. (22)**).

$$Fe^{2+} + HSO_5^- \rightarrow Fe^{3+} + SO_4^{\cdot -} + OH^- \quad (13)$$

$$Fe^{2+} + HSO_5^- \rightarrow Fe^{3+} + SO_4^{2-} + \cdot OH \quad (14)$$

$$Fe^{3+} + HSO_5^- \rightarrow Fe^{2+} + SO_5^{\cdot -} + H^+ \quad (15)$$

$$Mo^{4+} + 2\ HSO_5^- \rightarrow Mo^{6+} + 2\ SO_4^{\cdot -} + 2\ OH^- \quad (16)$$

$$Mo^{6+} + 2\ HSO_5^- \rightarrow Mo^{4+} + 2\ SO_5^{\cdot -} + 2\ H^+ \quad (17)$$

$$H_2O + SO_4^{\cdot -} \rightarrow \cdot OH + H^+ + SO_4^{2-} \quad (18)$$

$$2\ Fe^{3+} + Mo^{4+} \rightarrow Mo^{6+} + 2\ Fe^{2+} \quad (19)$$

$$Fe^{3+}/Mo^{6+} + S^{2-} \rightarrow Fe^{2+}/Mo^{4+} + \text{-}SO_n\text{-} \quad (20)$$

$$2SO_5^{\cdot -} + H_2O \rightarrow 2HSO_4^- + 1.5\ ^1O_2 \quad (21)$$

$$SMX + ROS\ (\cdot OH + ^1O_2 + SO_4^{\cdot -}) \rightarrow intermediates \rightarrow CO_2 + H_2O \quad (22)$$

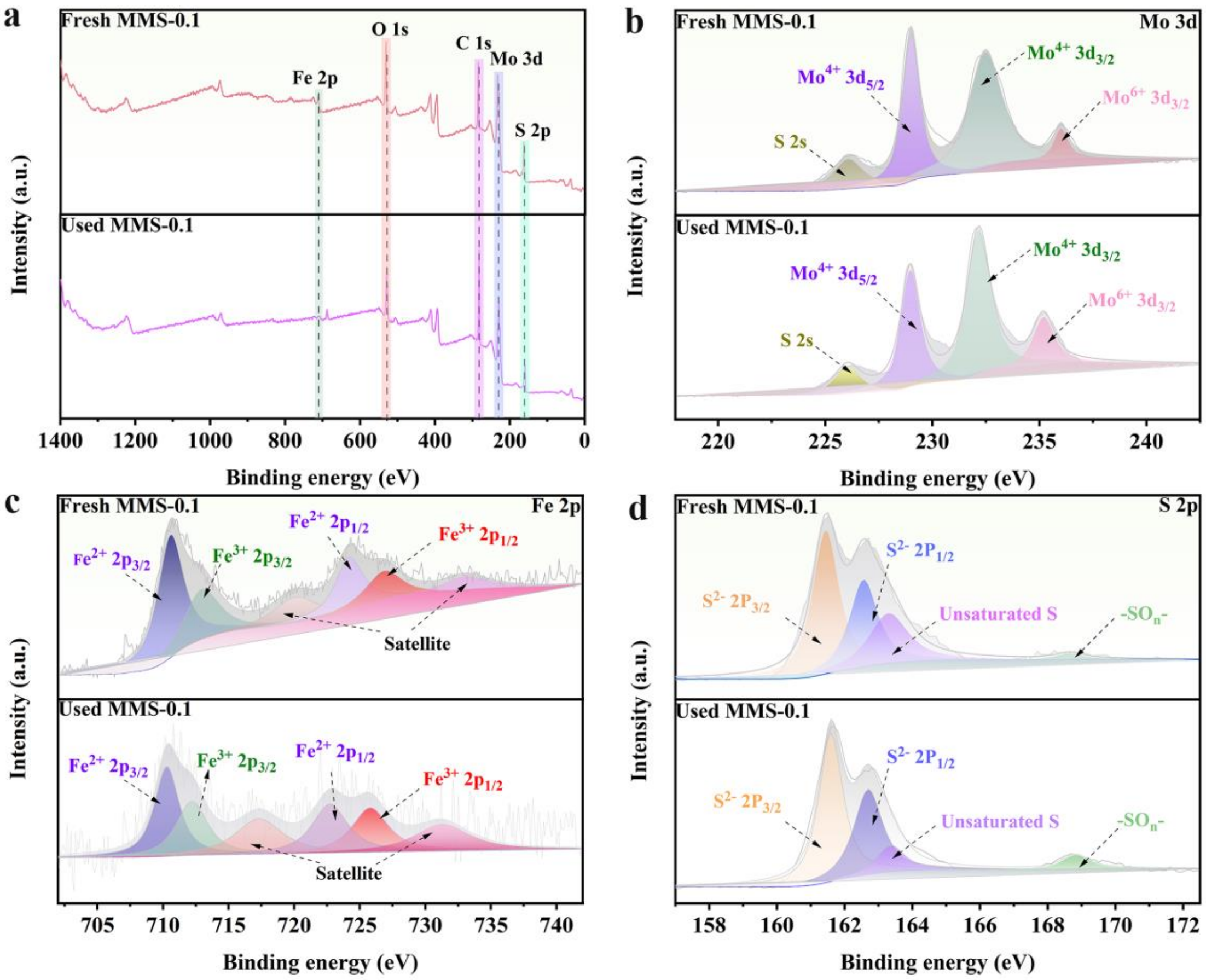

**Fig. 7.** The XPS survey spectra (a), high-resolution Mo 3d spectra (b), Fe 2p spectra (c), and S 2p spectra (d) of the fresh and used MMS-0.1.

To explore the existence of sulfur vacancies ($S_v$) in MMS-0.1, the low-temperature ESR spectra were studied [29, 36]. As shown in **Fig. S25a**, the signal peak of $S_v$ in MMS-0.1 was stronger than that of $MoS_2$, suggesting MMS-0.1 had more $S_v$ than $MoS_2$, which could be attributed to the introduction of MIL-101(Fe) on $MoS_2$ altered the content and distribution of $S_v$. It has been reported that $S_v$ could play several important roles in metallic sulfide-based catalysts but was not limited to follows: (i) $S_v$ could improve the electron density near the metal sites to accelerate electron transfer to enhance the electron-transfer ability of catalysts [62]. (ii) $S_v$ could expose more metal sites to enhance PMS activation [26]. Therefore, the electrochemical measurements including cyclic voltammetry (CV), electrochemical impedance spectroscopy (EIS), and linear sweep voltammetry (LSV) of catalysts were carried out to prove the role of $S_v$ in MMS-0.1 for PMS activation. As shown in **Fig. S25b,** MMS-0.1 showed a larger enclosed area than $MoS_2$ and MIL-101(Fe), indicating a higher reduction ability and better conductivity to coordinate the redox process, which contributed to accelerating electron transfer from catalyst to PMS. Moreover, EIS was further performed to corroborate the electron transfer capability of catalysts. As shown in **Fig. S25c,** the decreased semicircle size of MMS-0.1 also demonstrated that its conductivity was much better than that of $MoS_2$ and MIL-101(Fe). **Fig. S25d** showed the LSV of catalysts under a mixture of 500 mM $Na_2SO_4$ and 0.5 mM PMS. The distinct current increased in the MMS-0.1/PMS system than that of $MoS_2$/PMS and MIL-101(Fe)/PMS system. These above results suggested that MMS-0.1 with a large number of $S_v$ showed good conductivity and excellent electron transfer capability, which would be conducive to promoting $Fe^{2+}/Fe^{3+}$ cycle for enhancing PMS activation.

To further prove the contribution of $S_v$ in the activation of PMS by MMS-0.1, molecular simulation based on DFT was carried out at the molecular level **(Text S9)** [29, 63]. Two models were constructed, including MIL-101(Fe) and MMS-0.1 in **Fig. S26**. The PMS molecular adsorption onto the different active sites for the constructed models was simulated in **Fig. 8a**. Meanwhile, the corresponding bond lengths (lo-o) of $SO_3O$-OH (PMS molecules) and adsorption energies ($E_{ads}$) were summarized in **Table S8**. The adsorption energy of PMS on $\Delta E_{(Fe\text{-}3)}$, $\Delta E_{(Fe\text{-}1)}$ and $\Delta E_{(Mo\ and\ S_v)}$ was −2.878 eV, −2.545 eV and −1.511 eV, respectively. Obviously, the adsorption energy of PMS on Fe sites was higher than that of Mo and $S_v$ sites. In addition, $\Delta E_{(Fe\text{-}3)}$ and $\Delta E_{(Fe\text{-}1)}$ sites were more favorable on the prolonged $l_{O\text{-}O}$ bond length of PMS than that of $\Delta E_{(Mo\ and\ S_v)}$. The results confirmed that the PMS molecule could be effectively activated on the Fe sites of MMS-0.1 to produce ROS. It was well found that the adsorption energy of PMS onto the Fe sites of MMS-0.1 was greater than that of alone MIL-101(Fe). This indicated that the Mo sites and $S_v$ were conducive to promoting the PMS activation of Fe sites in MMS-0.1. **Fig. 8b-c** showed the differences in relative charge density after PMS adsorbed at MIL-101(Fe) and MMS-0.1. The charge accumulation region near PMS and the charge consumption region near Fe sites in MIL-101(Fe) and MMS-0.1 further proved the charge transfer from Fe sites to PMS during the activation process. Interestingly, it was found that the charge consumption area and charge accumulation area near the Fe sites in MMS-0.1 were larger compared with pure MIL-101(Fe), indicating that the presence of Mo sites and electron-rich $S_v$ in MMS-0.1 could accelerate the charge transfer from Fe sites to PMS. Therefore, the DFT results well revealed that the synergistic effect of $S_v$ and Mo sites accelerated the $Fe^{3+}/Fe^{2+}$ cycle to considerably improve the PMS activation. Based on the above data and interpretations, the degradation mechanism of SMX in the MMS-0.1/PMS system was proposed (**Fig. 8d**).

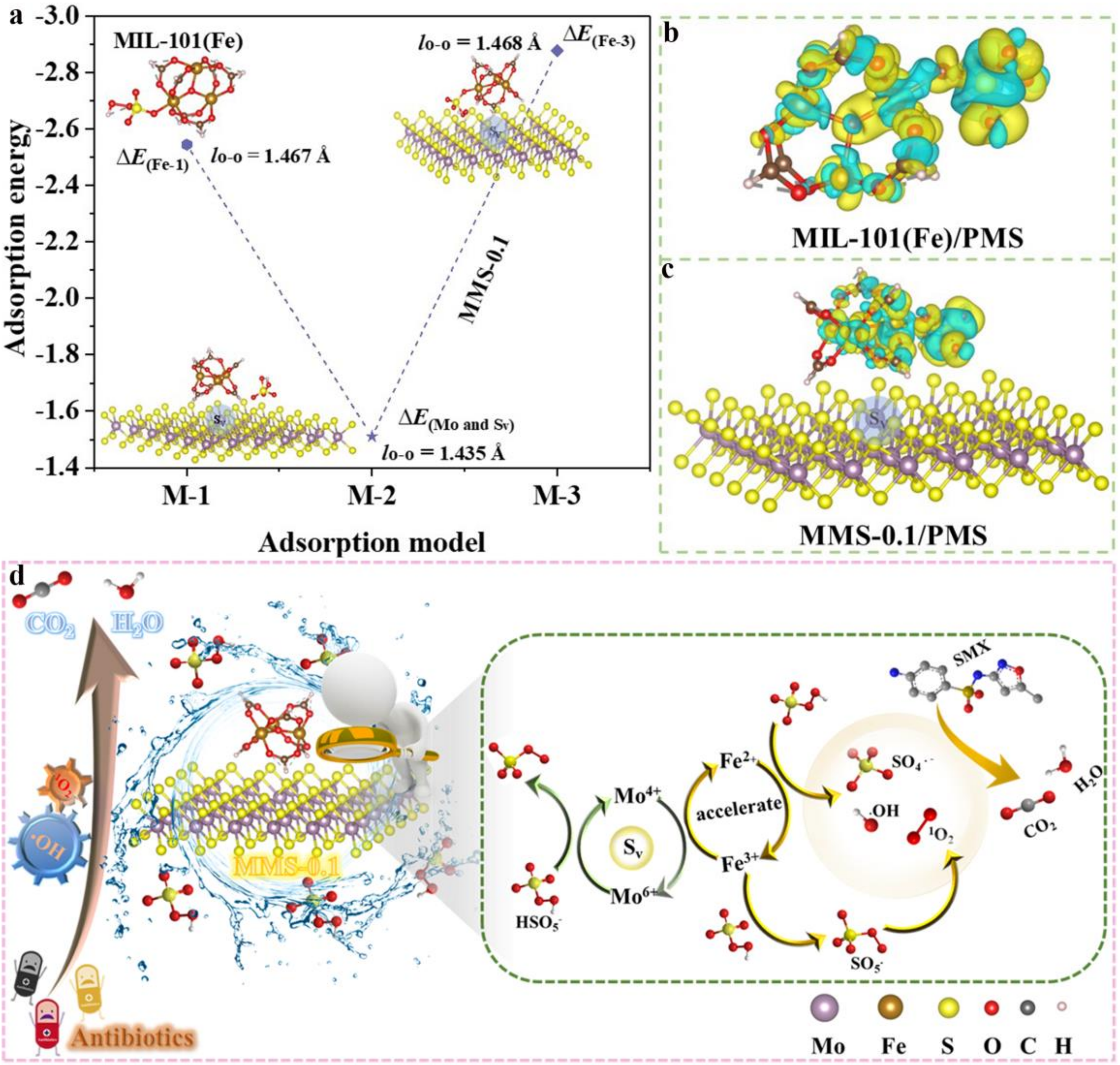

**Fig. 8.** (a) The adsorption energy results between the active sites and PMS in MIL-101(Fe) and MMS-0.1. The relative charge density differences after PMS adsorbed at (b) MIL-101(Fe) and (c) MMS-0.1 (the yellow and blue areas represent the charge aggregation and attenuation in space, respectively. the level of the iso-surface is set to 0.006 e Å-3). (d) Degradation mechanism of SMX in the MMS-0.1/PMS system.

### 3.6. Possible SMX degradation pathways and toxicity analysis

To better understand the degradation pathway of SMX, three-dimensional excitation-emission matrix fluorescence spectroscopy (3D EEMs) technology for SMX degradation in MMS-0.1/PMS/SMX system was investigated. In **Fig. S27a**, the fluorescence response at excitation wavelength/emission wavelength (Ex/Em) of 240-280/320-360 nm (region a) and Ex/Em of 330-350/380-410 nm (region b) was observed in the initial SMX solution (10 mg/L), belonging to tryptophan and fulvic acids, respectively, and the former is the main component [64]. As shown in **Fig. S27b**, after 2 min of degradation reaction, region b disappeared and the intensity of region a decreased slightly due to the low content of ROS in a short reaction time. When the reaction time reached 10 min, the fluorescence response of region a was weakened, but a new region for Ex/Em of 280-350/410-550 nm (region c) was observed in **Fig. S27c**. As the reaction time further increased to 15 min, the intensity of the region a and c reduced but not disappeared, indicating that the intermediate product of SMX was formed in the oxidation process (**Fig. S27d**). The above results were consistent with the mineralization rate of SMX (about 46%).

To predict the regional selectivity of SMX by ROS attack at the atomic level in MMS-0.1/PMS/SMX system, Multiwfn software was carried out to calculate the Fukui function ($f^0$) based on Hirshfeld charge, and the detailed descriptions were displayed in **Text S10** [65, 66]. Usually, the active sites obtained from the contour surfaces of $f^0$ for SMX with an isovalue of 0.01 (**Fig. S28a**) were consistent with the highest occupied molecular orbital (HOMO) (Calculated at B3LYP/6-31G* level; The red and green colors represent the positive and negative phases of the molecular orbitals, respectively.) distribution of SMX (**Fig. 9b**). Besides, the final structure optimized of SMX by Gaussian 09W software and the lowest occupied molecular orbital (LOMO) were shown in **Fig. 9a** and **Fig. 9c**. Although HOMO was vulnerable to ROS attack due to the area where electrons were lost, it cannot quantitatively depict reactivity at different sites. It was well-known that the reaction sites with high $f^0$ values indicated the site most vulnerable to ROS attack [32, 67]. Therefore, the calculated condensed $f^0$ of SMX was provided in **Fig. S28b**. It could be found that some sites including N17 ($f^0$= 0.1255), N2 ($f^0$= 0.0688), N10 ($f^0$= 0.0266), O1 ($f^0$= 0.0498), C5 ($f^0$= 0.056), C11 ($f^0$= 0.0471), and S7 ($f^0$= 0.0341) in the deprotonated SMX had a relatively good $f^0$, demonstrating that the sites were the more susceptible positions to be attacked by ROS. Based on the above analysis, the decomposition of SMX mainly occurs in N17, N2, N10, O1, C5, C11 and S7, resulting in the formation of bond cleavage C-N, isoxazole (N-O), S-N, C-O and C-S.

To further prove that SMX could be oxidated by ROS into small intermediate products, the degradation intermediates of SMX were monitored by LC-MS. **Fig. S29** depicted the mass spectra of the reaction intermediates of SMX degradation. In the meantime, the intermediate products with m/z of 242 (P1), 217 (P2), 246 (P3), 173 (P4), 338 (P5), 354 (P6), 98 (P7), 203 (P8), 158 (P9), 123 (P10), 102 (P11), 102 (P12), 58 (P13), and 85 (P14) were listed in **Table S9**. Based on mass spectral data and the Fukui function [68-70], the possible pathways for SMX degradation in the MMS-0.1/PMS system were proposed in **Fig. 9d**. Initially, the isoxazole ring of SMX was opened to the formation of P1 (m/z = 242). P1 could undergo hydroxylation to form P2 (m/z = 217) due to the attack of ·OH, and P2 were further oxidized by ROS to form P3 (m/z = 246). On the one hand, the S-N bond from SMX molecular (m/z = 254) was cleaved to generate P4 (m/z = 203) and P7 (m/z = 338) by ·OH, $SO_4^{\cdot-}$ and $^1O_2$. Then, P4 could undergo coupling reaction and oxidation to form P5 (m/z = 387) and P8 (m/z = 246), respectively. Subsequently, the C-S and C-N bonds of P8 were attacked by ROS to produce P10 (m/z = 123) and P9 (m/z = 158), respectively. P7 could be simultaneously transferred into P11 (m/z = 102) and P12 (m/z = 102) by ring-opening and hydroxylation reaction. After that, P11 and P12 were further oxidized and ring-opening reactions into P13 (m/z = 58) and P14 (m/z = 85). Ultimately, the above-generated intermediates would be mineralized into harmless $CO_2$, $H_2O$ and inorganic ions.

The toxicity of the intermediates of SMX degradation was predicted by the Toxicity Estimation Software (T.E.S.T.) with QSAR analysis [71]. As shown in **Fig. 9e**, except for P10 (m/z = 123), the bioaccumulation factor of the intermediates was lower than that of SMX, suggesting the bioaccumulation significantly decreased in the MMS-0.1/PMS/SMX system. In addition, **Fig. 9f** showed that both SMX and its intermediates both were mutagenicity-negative products. However, it was found that SMX was "developmental toxicant" in **Fig. 9g**. Fortunately, most of the intermediates of SMX for developmental toxicity were reduced, and even some intermediates were nontoxic products. Although a few intermediates were still low-toxic, the overall toxicity of these intermediates could be reduced by prolonging the reaction time due to the good mineralization ability of the system (**Table S10**). Therefore, the MMS-0.1/PMS system could not only effectively degrade SMX, but also obtain the nontoxic transformation of SMX, which was expected to be used for actual wastewater purification.

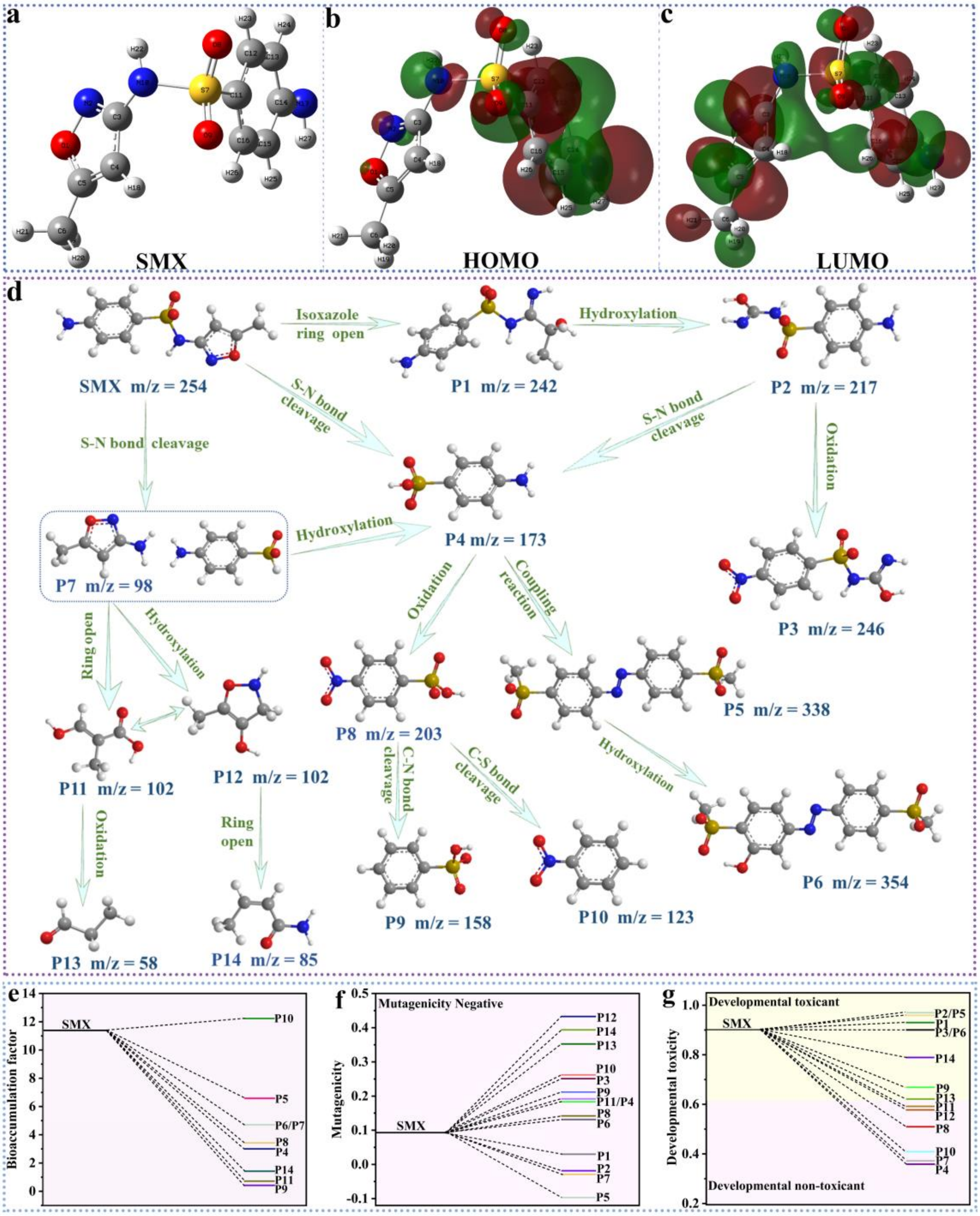

**Fig. 9.** (a) Chemical structure of SMX after optimization (gray: C; white: H; blue: N; red: O; yellow: S). (b) HOMO and (c) LUMO orbital based on the isosurface of Fukui

index. (d) The proposed degradation pathway of SMX in the MMS-0.1/PMS system. (e) Bioaccumulation factor, (f) mutagenicity and (g) developmental toxicity of SMX of the degradation intermediates.

**4. Conclusion**

In summary, we have successfully constructed garland-like MIL-101(Fe)/$MoS_2$ nanosheets for the direct activation of PMS to sustainably remove SMX. In comparison with the single counterparts and the previously reported candidates, the superior performance of the optimized MMS-0.1 was due to the following striking features: (i) The unique garland-like structure could expose more reaction sites to efficiently promote the rapid internal electron transfer between MIL-101(Fe) and $MoS_2$. (ii) Rich $S_v$ and $Mo^{6+}/Mo^{4+}$ sites could accelerate the $Fe^{3+}/Fe^{2+}$ cycle in the PMS activation process. (iii) Synergistic of radical ($\cdot OH$) and non-radical ($^1O_2$) mechanism was beneficial for the highly efficient degradation of SMX. The MMS-0.1/PMS system showed high SMX degradation efficiency of 100% and PMS utilization rate of 90% within 30 min, which possessed the mineralization efficiency of 46.2%. Besides, the MMS-0.1 catalyst exhibited excellent stability and reusability, where it could still reach 92% of the SMX degradation rate without extra regeneration treatment after five cycles. The nontoxic transformation of SMX could be realized by attacking the C-N, N-O, S-N, C-O and C-S bonds of SMX using ROS in the MMS-0.1/PMS/SMX system. This work paves the way for the rational design of Fe-$MoS_2$ nanosheets with dual metal active sites and rich sulfur vacancies for the sustainable degradation of organic pollutants.

**Acknowledgements**

This work is financially supported by National Natural Science Foundation of China (22078374, 21776324), the National Ten Thousand Talent Plan, Key Realm Research and Development Program of Guangdong Province (2020B0202080001)，the Guangdong Basic and Applied Basic Research Foundation (2019B1515120058), Science and Technology Planning Project of Guangdong Province, China (2021B1212040008), Guangdong Laboratory for Lingnan Modern Agriculture Project (NT2021010), the Scientific and Technological Planning Project of Guangzhou (202206010145).

**References**

[1] M. Patel, R. Kumar, K. Kishor, T. Mlsna, C.U. Pittman, D. Mohan, Pharmaceuticals of emerging concern in aquatic systems: Chemistry, occurrence, effects, and removal methods, Chem. Rev. 119(6) (2019) 3510-3673.

[2] D. KENNEDY, Time to deal with antibiotics, Science 342(6160) (2013) 777-777.

[3] J. Li, L. Zhao, M. Feng, C.-H. Huang, P. Sun, Abiotic transformation and ecotoxicity change of sulfonamide antibiotics in environmental and water treatment processes: A critical review, Water Res. 202 (2021) 117463.

[4] P.J. Vikesland, A. Pruden, P.J.J. Alvarez, D. Aga, H. Burgmann, X.D. Li, C.M. Manaia, I. Nambi, K. Wigginton, T. Zhang, Y.G. Zhu, Toward a comprehensive strategy to mitigate dissemination of environmental sources of antibiotic resistance, Environ. Sci. Technol. 51(22) (2017) 13061-13069.

[5] L. Qin, W. Chen, Y. Fu, J. Tang, H. Yi, L. Li, F. Xu, M. Zhang, W. Cao, D. Huang, C. Lai, Hemin derived iron and nitrogen-doped carbon as a novel heterogeneous electro-Fenton catalyst to efficiently degrade ciprofloxacin, Chem. Eng. J. 449 (2022) 137840.

[6] M. Li, S. You, X. Duan, Y. Liu, Selective formation of reactive oxygen species in peroxymonosulfate activation by metal-organic framework-derived membranes: A defect engineering-dependent study, Appl. Catal., B 312 (2022) 121419.

[7] X. Zhang, B. Xu, S. Wang, X. Li, C. Wang, B. Liu, F. Han, Y. Xu, P. Yu, Y. Sun, Tetracycline degradation by peroxymonosulfate activated with $CoN_x$ active sites: Performance and activation mechanism, Chem. Eng. J. 431 (2022) 133477.

[8] L. Qin, H. Ye, C. Lai, S. Liu, X. Zhou, F. Qin, D. Ma, B. Long, Y. Sun, L. Tang, M. Yan, W. Chen, W. Chen, L. Xiang, Citrate-regulated synthesis of hydrotalcite-like compounds as peroxymonosulfate activator-Investigation of oxygen vacancies and degradation pathways by combining DFT, Appl. Catal., B 317 (2022) 121704.

[9] Z. Guo, C. Li, M. Gao, Xi. Han, Y. Zhang, W. Zhang, W. Li, Mn@OCovalency governs the intrinsic activity of Co-Mn spinel oxides for boosted peroxymonosulfate activation, Angew. Chem. Int. Ed. 60 (2021) 274-280.

[10] W. Li, S. Xia, Z. Wang, B. Zhang, B. Li, L. Zhang, K. Qian, J. Ma, X. He, Covalency competition triggers Fe-Co synergistic catalysis for boosted Fenton-like reactions, Appl. Catal., B 325 (2023) 122358.

[11] S. Zhou, X. Miao, X. Zhao, C. Ma, Y. Qiu, Z. Hu, J. Zhao, L. Shi, J. Zeng, Engineering electrocatalytic activity in nanosized perovskite cobaltite through surface spin-state transition, Nat. Commun. 7(1) (2016) 11510.

[12] J. Xu, X. Zheng, Z. Feng, Z. Lu, Z. Zhang, W. Huang, Y. Li, D. Vuckovic, Y. Li, S. Dai, G. Chen, K. Wang, H. Wang, J.K. Chen, W. Mitch, Y. Cui, Organic wastewater treatment by a single-atom catalyst and electrolytically produced $H_2O_2$, Nat. Sustain. 4 (2021) 233-241.

[13] D. Meyerstein, Re-examining Fenton and Fenton-like reactions, Nat. Rev. Chem. 5(9) (2021) 595-597.

[14] F. Chen, X.L. Wu, C. Shi, H. Lin, J. Chen, Y. Shi, S. Wang, X. Duan, Molecular engineering toward pyrrolic N-rich M-$N_4$ (M = Cr, Mn, Fe, Co, Cu) single-atom sites for enhanced heterogeneous Fenton-Like reaction, Adv. Funct. Mater. 31(13) (2021) 2007877.

[15] R. Zhu, Y. Zhu, H. Xian, L. Yan, H. Fu, G. Zhu, Y. Xi, J. Zhu, H. He, CNTs/ferrihydrite as a highly efficient heterogeneous Fenton catalyst for the degradation of bisphenol A: The important role of CNTs in accelerating Fe(III)/Fe(II) cycling, Appl. Catal., B 270 (2020) 118891.

[16] X. Li, X. Huang, S. Xi, S. Miao, J. Ding, W. Cai, S. Liu, X. Yang, H. Yang, J. Gao, J. Wang, Y. Huang, T. Zhang, B. Liu, Single cobalt atoms anchored on porous N-doped graphene with dual reaction sites for efficient Fenton-like catalysis, J. Am. Chem. Soc. 140(39) (2018) 12469-12475.

[17] G. Liu, X. Li, B. Han, L. Chen, L. Zhu, L.C. Campos, Efficient degradation of sulfamethoxazole by the Fe(II)/$HSO_5^-$ process enhanced by hydroxylamine: Efficiency and mechanism, J. Hazard. Mater. 322(Pt B) (2017) 461-468.

[18] J. Zou, J. Ma, L. Chen, X. Li, Y. Guan, P. Xie, C. Pan, Rapid acceleration of ferrous iron/peroxymonosulfate oxidation of organic pollutants by promoting Fe(III)/Fe(II) cycle with hydroxylamine, Environ. Sci. Technol. 47(20) (2013) 11685-91.

[19] A. Rastogi, S.R. Al-Abed, D.D. Dionysiou, Effect of inorganic, synthetic and naturally occurring chelating agents on Fe(II) mediated advanced oxidation of chlorophenols, Water Res. 43(3) (2009) 684-94.

[20] T. Li, Z. Zhao, Q. Wang, P. Xie, J. Ma, Strongly enhanced Fenton degradation of organic pollutants by cysteine: An aliphatic amino acid accelerator outweighs hydroquinone analogues, Water Res. 105 (2016) 479-486.

[21] L. Bu, C. Bi, Z. Shi, S. Zhou, Significant enhancement on ferrous/persulfate oxidation with epigallocatechin-3-gallate: Simultaneous chelating and reducing, Chem. Eng. J. 321 (2017) 642-650.

[22] M. Huang, X. Wang, C. Liu, G. Fang, J. Gao, Y. Wang, D. Zhou, Mechanism of metal sulfides accelerating Fe(II)/Fe(III) redox cycling to enhance pollutant degradation by persulfate: Metallic active sites vs. reducing sulfur species, J. Hazard. Mater. 404(Pt B) (2021) 124175.

[23] M. Cai, R. Li, Z. Xie, J. Huang, Y. Zeng, Q. Zhang, H. Liu, W. Lv, G. Liu, Synthesis of a core-shell heterostructured $MoS_2/Cd_{0.9}Zn_{0.1}S$ photocatalyst for the degradation of diclofenac under visible light, Appl. Catal., B 259 (2019) 118033.

[24] L. Zhu, J. Ji, J. Liu, S. Mine, M. Matsuoka, J. Zhang, M. Xing, Designing 3D-$MoS_2$ sponge as excellent cocatalysts in advanced oxidation processes for pollutant control, Angew. Chem. Int. Ed. 59 (2022) 13968-13976.

[25] C. Xiao, Y. Hu, Q. Li, J. Liu, X. Li, Y. Shi, Y. Chen, J. Cheng, Carbon-doped defect $MoS_2$ co-catalytic $Fe^{3+}$/peroxymonosulfate process for efficient sulfadiazine degradation: Accelerating $Fe^{3+}/Fe^{2+}$ cycle and $^1O_2$ dominated oxidation, Sci. Total. Environ. 858(Pt 1) (2023) 159587.

[26] B. Sheng, F. Yang, Y. Wang, Z. Wang, Q. Li, Y. Guo, X. Lou, J. Liu, Pivotal roles of $MoS_2$ in boosting catalytic degradation of aqueous organic pollutants by Fe(II)/PMS, Chem. Eng. J. 375 (2019) 121989.

[27] S. Wang, W. Xu, J. Wu, Q. Gong, P. Xie, Improved sulfamethoxazole degradation by the addition of $MoS_2$ into the $Fe^{2+}$/peroxymonosulfate process, Sep. Purif. Technol. 235 (2020) 116170.

[28] L.-Z. Huang, X. Wei, E. Gao, C. Zhang, X.-M. Hu, Y. Chen, Z. Liu, N. Finck, J. Lützenkirchen, D.D. Dionysiou, Single Fe atoms confined in two-dimensional $MoS_2$ for sulfite activation: A biomimetic approach towards efficient radical generation, Appl. Catal., B 268 (2020) 118459.

[29] X. Li, L. Wang, Y. Guo, W. Song, Y. Li, L. Yan, Goethite-$MoS_2$ hybrid with dual active sites boosted peroxymonosulfate activation for removal of tetracycline:

The vital roles of hydroxyl radicals and singlet oxygen, Chem. Eng. J. 450 (2022) 138104.

[30] J. Lu, Y. Zhou, Y. Zhou, Efficiently activate peroxymonosulfate by $Fe_3O_4$@$MoS_2$ for rapid degradation of sulfonamides, Chem. Eng. J. 422 (2021) 130126.

[31] F. Zhao, Y. Liu, S.B. Hammouda, B. Doshi, N. Guijarro, X. Min, C.-J. Tang, M. Sillanpää, K. Sivula, S. Wang, MIL-101(Fe)/g-$C_3N_4$ for enhanced visible-light-driven photocatalysis toward simultaneous reduction of Cr(VI) and oxidation of bisphenol A in aqueous media, Appl. Catal., B 272 (2020) 119033.

[32] F. Wang, S.S. Liu, Z. Feng, H. Fu, M. Wang, P. Wang, W. Liu, C.C. Wang, High-efficient peroxymonosulfate activation for rapid atrazine degradation by $FeS_x$@$MoS_2$ derived from MIL-88A(Fe), J. Hazard. Mater. 440 (2022) 129723.

[33] C. Vallés-García, E. Gkaniatsou, A. Santiago-Portillo, M. Giménez-Marqués, M. Álvaro, J.-M. Greneche, N. Steunou, C. Sicard, S. Navalón, C. Serre, H. García, Design of stable mixed-metal MIL-101(Cr/Fe) materials with enhanced catalytic activity for the Prins reaction, J. Mater. Chem. A 8(33) (2020) 17002-17011.

[34] J.-C.E. Yang, M.-P. Zhu, X. Duan, S. Wang, B. Yuan, M.-L. Fu, The mechanistic difference of 1T-2H $MoS_2$ homojunctions in persulfates activation: Structure-dependent oxidation pathways, Appl. Catal., B 297 (2021) 120460.

[35] L. Cai, J. He, Q. Liu, T. Yao, L. Chen, W. Yan, F. Hu, Y. Jiang, Y. Zhao, T. Hu, Z. Sun, S. Wei, Vacancy-induced ferromagnetism of $MoS_2$ nanosheets, J. Am. Chem. Soc. 137(7) (2015) 2622-7.

[36] R. Bai, W. Yan, Y. Xiao, S. Wang, X. Tian, J. Li, X. Xiao, X. Lu, F. Zhao, Acceleration of peroxymonosulfate decomposition by a magnetic $MoS_2/CuFe_2O_4$ heterogeneous catalyst for rapid degradation of fluoxetine, Chem. Eng. J. 397 (2020) 125501.

[37] H. Akram, C. Mateos-Pedrero, E. Gallegos-Suárez, A. Guerrero-Ruíz, T. Chafik, I. Rodríguez-Ramos, Effect of electrolytes nature and concentration on the morphology and structure of $MoS_2$ nanomaterials prepared using one-pot solvothermal method, Appl. Surf. Sci. 307 (2014) 319-326.

[38] X. Duan, H. Sun, S. Wang, Metal-Free Carbocatalysis in Advanced Oxidation Reactions, Acc. Chem. Res 51 (2018) 678-687.

[39] J. Zhou, X. Guo, X. Zhou, J. Yang, S. Yu, X. Niu, Q. Chen, F. Li, Y. Liu, Boosting the efficiency of Fe-$MoS_2$/peroxymonosulfate catalytic systems for organic powllutants remediation: Insights into edge-site atomic coordination, Chem. Eng. J. 433 (2022) 134511.

[40] H. Fu, S. Ma, P. Zhao, S. Xu, S. Zhan, Activation of peroxymonosulfate by graphitized hierarchical porous biochar and $MnFe_2O_4$ magnetic nanoarchitecture for organic pollutants degradation: Structure dependence and mechanism, Chem. Eng. J. 360 (2019) 157-170.

[41] Y. Gao, Y. Zhu, T. Li, Z. Chen, Q. Jiang, Z. Zhao, X. Liang, C. Hu, Unraveling the High-Activity Origin of Single-Atom Iron Catalysts for Organic Pollutant Oxidation via Peroxymonosulfate Activation, Environ. Sci. Technol. 55(12) (2021) 8318-8328.

[42] X. Li, K. Hu, Y. Huang, Q. Gu, Y. Chen, B. Yang, R. Qiu, W. Luo, B.M. Weckhuysen, K. Yan, Upcycling biomass waste into Fe single atom catalysts for pollutant control, J. Energy Chem. 69 (2022) 282-291.

[43] Y. Long, Y. Huang, H. Wu, X. Shi, L. Xiao, Peroxymonosulfate activation for pollutants degradation by Fe-N-codoped carbonaceous catalyst: Structure-dependent performance and mechanism insight, Chem. Eng. J. 369 (2019) 542-552.

[44] N. Song, S. Ren, Y. Zhang, C. Wang, X. Lu, Confinement of prussian blue analogs boxes inside conducting polymer nanotubes enables significantly enhanced catalytic performance for water treatment, Adv. Funct. Mater. 32(34) (2022) 2204751.

[45] X. Ao, W. Sun, S. Li, C. Yang, C. Li, Z. Lu, Degradation of tetracycline by medium pressure UV-activated peroxymonosulfate process: Influencing factors, degradation pathways, and toxicity evaluation, Chem. Eng. J. 361 (2019) 1053-1062.

[46] L. Peng, X. Duan, Y. Shang, B. Gao, X. Xu, Engineered carbon supported single iron atom sites and iron clusters from Fe-rich Enteromorpha for Fenton-like reactions via nonradical pathways, Appl. Catal., B 287 (2021) 119963.

[47] P. Cai, J. Zhao, X. Zhang, T. Zhang, G. Yin, S. Chen, C.-L. Dong, Y.-C. Huang, Y. Sun, D. Yang, B. Xing, Synergy between cobalt and nickel on $NiCo_2O_4$ nanosheets promotes peroxymonosulfate activation for efficient norfloxacin degradation, Appl. Catal., B 306 (2022) 121091.

[48] A. Jawad, K. Zhan, H. Wang, A. Shahzad, Z. Zeng, J. Wang, X. Zhou, H. Ullah, Z. Chen, Z. Chen, Tuning of persulfate activation from a free radical to a nonradical pathway through the incorporation of non-redox magnesium oxide, Environ. Sci. Technol. 54(4) (2020) 2476-2488.

[49] W. Ma, N. Wang, Y. Fan, T. Tong, X. Han, Y. Du, Non-radical-dominated catalytic degradation of bisphenol A by ZIF-67 derived nitrogen-doped carbon nanotubes frameworks in the presence of peroxymonosulfate, Chem. Eng. J. 336 (2018) 721-731.

[50] Y. Mei, Y. Qi, J. Li, X. Deng, S. Ma, T. Yao, J. Wu, Construction of yolk/shell $Fe_3O_4$@$MgSiO_3$ nanoreactor for enhanced Fenton-like reaction via spatial separation of adsorption sites and activation sites, J. Taiwan Inst. Chem. Eng. 113 (2020) 363-371.

[51] G. Fang, J. Gao, D.D. Dionysiou, C. Liu, D. Zhou, Activation of persulfate by quinones: free radical reactions and implication for the degradation of PCBs, Environ. Sci. Technol. 47(9) (2013) 4605-11.

[52] J. Yang, M. Zhang, M. Chen, Y. Zhou, M. Zhu, Oxygen vacancies in piezoelectric ZnO twin-mesocrystal to improve peroxymonosulfate utilization efficiency via piezo-activation for antibiotic ornidazole removal, Adv. Mater. (2023) 2209885.

[53] P. Shao, J. Tian, F. Yang, X. Duan, S. Gao, W. Shi, X. Luo, F. Cui, S. Luo, S. Wang, Identification and regulation of active sites on nanodiamonds: establishing

a highly efficient catalytic system for oxidation of organic contaminants, Adv. Funct. Mater. 28(13) (2018) 1705295.

[54] X. Chen, W.-D. Oh, T.-T. Lim, Graphene- and CNTs-based carbocatalysts in persulfates activation: Material design and catalytic mechanisms, Chem. Eng. J. 354 (2018) 941-976.

[55] P. Shao, Y. Jing, X. Duan, H. Lin, L. Yang, W. Ren, F. Deng, B. Li, X. Luo, S. Wang, Revisiting the graphitized nanodiamond-mediated activation of peroxymonosulfate: Singlet oxygenation versus electron transfer, Environ. Sci. Technol. 55(23) (2021) 16078-16087.

[56] L. Jin, S. You, N. Ren, B. Ding, Y. Liu, Mo Vacancy-Mediated Activation of peroxymonosulfate for ultrafast micropollutant removal using an electrified MXenefilter functionalized with Fe single atoms, Environ. Sci. Technol. 56(16) (2022) 11750-11759.

[57] X. Li, S. Wang, P. Chen, B. Xu, X. Zhang, Y. Xu, R. Zhou, Y. Yu, H. Zheng, P. Yu, Y. Sun, ZIF-derived non-bonding Co/Zn coordinated hollow carbon nitride for enhanced removal of antibiotic contaminants by peroxymonosulfate activation: Performance and mechanism, Appl. Catal., B 325 (2023) 122401.

[58] Z. Wang, E. Almatrafi, H. Wang, H. Qin, W. Wang, L. Du, S. Chen, G. Zeng, P. Xu, Cobalt single atoms anchored on oxygen-doped tubular carbon nitride for efficient peroxymonosulfate activation: Simultaneous coordination structure and morphology modulation, Angew. Chem. Int. Ed. 61 (2022) e202202338.

[59] P. Shao, S. Yu, X. Duan, L. Yang, H. Shi, L. Ding, J. Tian, L. Yang, X. Luo, S. Wang, Potential difference driving electron transfer via defective carbon nanotubes toward selective oxidation of organic micropollutants, Environ. Sci. Technol. 54(13) (2020) 8464-8472.

[60] X. Wu, W. Zhao, Y. Huang, G. Zhang, A mechanistic study of amorphous $CoS_x$ cages as advanced oxidation catalysts for excellent peroxymonosulfate activation towards antibiotics degradation, Chem. Eng. J. 381 (2020) 122768.

[61] C. Zhou, L. Zhu, L. Deng, H. Zhang, H. Zeng, Z. Shi, Efficient activation of peroxymonosulfate on CuS@MIL-101(Fe) spheres featured with abundant sulfur vacancies for coumarin degradation: Performance and mechanisms, Sep. Purif. Technol. 276 (2021) 119404.

[62] H. Kuang, Z. He, M. Li, R. Huang, Y. Zhang, X. Xu, L. Wang, Y. Chen, S. Zhao, Enhancing co-catalysis of $MoS_2$ for persulfate activation in $Fe^{3+}$-based advanced oxidation processes via defect engineering, Chem. Eng. J. 417 (2021) 127987.

[63] C. Bao, H. Wang, C. Wang, X. Zhang, X. Zhao, C.-L. Dong, Y.-C. Huang, S. Chen, P. Guo, X. She, Y. Sun, D. Yang, Cooperation of oxygen vacancy and $Fe^{III}/Fe^{II}$ sites in $H_2$-reduced Fe-MIL-101 for enhanced Fenton-like degradation of organic pollutants, J. Hazard. Mater. 441 (2023) 129922.

[64] W. Li, J. Zhu, Y. Lou, A. Fang, H. Zhou, B. Liu, G. Xie, D. Xing, $MnO_2$/tourmaline composites as efficient cathodic catalysts enhance bioelectroremediation of contaminated river sediment and shape biofilm

microbiomes in sediment microbial fuel cells, Appl. Catal., B 278 (2020) 119331.

[65] W. Qin, Z. Lin, H. Dong, X. Yuan, Z. Qiang, S. Liu, D. Xia, Kinetic and mechanistic insights into the abatement of clofibric acid by integrated UV/ozone/peroxydisulfate process: A modeling and theoretical study, Water Res 186 (2020) 116336.

[66] T. Lu, F. Chen, Multiwfn: a multifunctional wavefunction analyzer, J. Comput. Chem. 33(5) (2012) 580-92.

[67] Z. Cai, X. Hao, X. Sun, P. Du, W. Liu, J. Fu, Highly active $WO_3$@anatase-$SiO_2$ aerogel for solar-light-driven phenanthrene degradation: Mechanism insight and toxicity assessment, Water Res. 162 (2019) 369-382.

[68] H. Milh, D. Cabooter, R. Dewil, Role of process parameters in the degradation of sulfamethoxazole by heat-activated peroxymonosulfate oxidation: Radical identification and elucidation of the degradation mechanism, Chem. Eng. J. 422 (2021) 130457.

[69] A. Wang, J. Ni, W. Wang, D. Liu, Q. Zhu, B. Xue, C.-C. Chang, J. Ma, Y. Zhao, MOF Derived Co−Fe nitrogen doped graphite carbon@crosslinked magnetic chitosan Micro−nanoreactor for environmental applications: Synergy enhancement effect of adsorption−PMS activation, Appl. Catal., B 319 (2022) 121926.

[70] S. Wang, Y. Liu, J. Wang, Peroxymonosulfate Activation by Fe-Co-O-Codoped Graphite Carbon Nitride for Degradation of Sulfamethoxazole, Environ. Sci. Technol. 54(16) (2020) 10361-10369.

[71] X. Long, Z. Xiong, R. Huang, Y. Yu, P. Zhou, H. Zhang, G. Yao, B. Lai, Sustainable Fe(III)/Fe(II) cycles triggered by co-catalyst of weak electrical current in Fe(III)/peroxymonosulfate system: Collaboration of radical and non-radical mechanisms, Appl. Catal., B 317 (2022) 121716.